\documentclass[fleqn,usenatbib]{mnras}

\usepackage{newtxtext,newtxmath}
\usepackage{comment}
\usepackage{array}
\usepackage{float}
\usepackage{stfloats}
\usepackage{tabularray}
\usepackage{tabularx}
\usepackage{threeparttable}
\UseTblrLibrary{booktabs}
\usepackage{lipsum}
\usepackage{multirow}
\usepackage{mathtools}
\usepackage{adjustbox}

\usepackage[T1]{fontenc}
\usepackage{ae,aecompl}
\usepackage{graphicx}	
\usepackage{amsmath}	
\usepackage{enumitem} 
\usepackage{xfrac} 
\usepackage{algorithm}
\usepackage{algpseudocode}
\usepackage{geometry}

\setlist[itemize]{leftmargin=*}

\newcommand{\mstar}{\texttt{MSTAR}}

\newcommand{\bifrost}{\texttt{BIFROST}}

\usepackage{hyperref}
\hypersetup{
    colorlinks=blue,
    linkcolor=blue,
    citecolor=blue,
    urlcolor=blue
}

\title[Plunges and IMRIs from Single and Binary SMBHs]{Elusive Plunges and Heavy Intermediate-mass-ratio Inspirals from Single and Binary Supermassive Black Holes}

\author[Souvaitzis et al.]{
Lazaros Souvaitzis$^{1}$\thanks{E-mail: lazaros@mpa-garching.mpg.de}, 
Antti Rantala$^{1}$
and Thorsten Naab$^{1}$ 
\\
$^{1}$Max Planck Institute for Astrophysics, Karl-Schwarzschild-Str. 1, 85748, Garching, Germany\\
}

\date{Accepted XXX. Received YYY; in original form ZZZ}

\pubyear{2025}

\begin{document}
\label{firstpage}
\pagerange{\pageref{firstpage}--\pageref{lastpage}}
\maketitle


\begin{abstract}
The most massive galaxies in the Universe also host the largest supermassive black holes (SMBHs), with masses of $10^9 \: \mathrm{M_{\odot}}$ and above. During their hierarchical assembly, these galaxies have experienced only a few major mergers at low redshift, but have accreted many low-mass galaxies across cosmic time, possibly hosting intermediate mass black holes (IMBHs). If some of these IMBHs migrate to the galactic center, they may form compact subsystems around the central SMBH. We investigate the evolution of such subsystems, consisting of ten $10^5 \: \mathrm{M_{\odot}}$ IMBHs at three different concentrations around a $10^9 \: \mathrm{M_{\odot}}$ SMBH. We evolve these systems both in isolation and in the presence of a companion SMBH, using \texttt{MSTAR}, a regularized integration method including relativistic effects up to post-Newtonian order 3.5PN. Our analysis focuses on gravitational-wave–driven intermediate-mass-ratio inspirals (heavy IMRIs) and direct plunges. We show that perturbations from a secondary SMBH enhance the number of IMBH direct plunges by more than a factor of two, making them the dominant merger channel. These plunges and IMRIs with a central $10^9 \: \mathrm{M_{\odot}}$ SMBH will contribute to SMBH growth but will likely evade detection with future gravitational-wave interferometers and pulsar timing arrays (PTAs). However, for galaxies with lower-mass SMBHs ($M_\bullet \lesssim 10^8 \:\mathrm{M_{\odot}}$), heavy IMRIs will be detectable with the Laser Interferometer Space Antenna (LISA) and can provide direct observational constraints on the existence of IMBHs, while the more numerous plunges will still remain hidden.
\end{abstract}



\section{Introduction}

Supermassive black holes (SMBHs) are expected to be located at the centers of practically all massive galaxies, with masses spanning approximately $10^6-10^{10} \: \mathrm{M_{\odot}}$ \citep{Kormendy_1995,Menci_2004,Kormendy_2013}. Within the framework of the $\mathrm{\Lambda CDM}$ paradigm of structure formation, galaxies assemble hierarchically: the systems form small and subsequently grow and merge to produce more massive ones \citep{Ostriker_1977,White_1980,Lacey_1993}. If the galaxies contain massive black holes, galactic mergers naturally lead to the creation of SMBH pairs, with the two black holes initially separated by $\mathrm{kpc}$ distances within the merger remnant \citep{Barnes_1992}. Their subsequent orbital evolution is driven by a range of dynamical processes, that remove energy and angular momentum from the pair, bringing them closer together and potentially culminating in their coalescence \citep{Begelman_1980Natur}. Understanding the origin and subsequent growth of such SMBHs remains one the fundamental research topics in modern astrophysics \citep{Rees_1988,Volonteri_2021}.

\emph{Major} galaxy mergers—interactions between galaxies of comparable mass—play a key role in the assembly of massive galaxies and have been extensively studied out to high redshift \citep{Conselice_2003,Bluck_2009,Bluck_2012,Lotz_2011,Man_2016,Mundy_2017,Mantha_2018,Duncan_2019,Ferreira_2020,Dai_2021,Husko_2022,Shibuya_2022}. Their dynamical and morphological evolution, also including their supermassive black holes, has been investigated with numerical simulations \citep[e.g.,][]{Mihos_1996,Hopkins_2008,Johansson_2009,Dekel_2009,Hilz_2012,Pillepich_2018,Hani_2018,Rantala2018,Mannerkoski2021}. Mergers with more unequal mass ratios like \emph{minor} mergers with mass ratios of $4 \lesssim M_1/M_2 < 10$, \emph{very minor} or \emph{mini} mergers with mass-ratios $10 \lesssim M_1/M_2 < 100$, and so-called accretion events $M_1/M_2 > 100$ are much more common \citep[e.g.][]{Nipoti_2025}. Semi-analytical, empirical, and numerical cosmological models all agree that the most massive early-type galaxies these merger and accretion events dominate the mass assembly of massive early-type galaxies, in particular at low redshift \citep[e.g.][]{De_Lucia_2007,Moster_2013,2013MNRAS.435..901L,Rodriguez_Gomez_2016,Moster_2018,Behroozi_2019,Moster_2020} such that most of their stellar mass is of ex-situ origin. There is theoretical evidence that the accreted stellar mass mixes well in case of major mergers and the central supermassive black holes sink to the center and eventually merge \citep[e.g.][]{Springel_2005,Di_Matteo_2008}. For minor mergers and accretion events, the accreted stars assemble at larger radii \citep[e.g.][]{2013MNRAS.429.2924H} which is supported by observations as well as simulations \citep[e.g.][]{van_Dokkum_2010,Hilz_2012,D_Souza_2014,Williams_2025}. Additional observational evidence for galaxy growth through mergers up to redshift $z < 3$ has been reported by \cite{Conselice_2022}, who find that galaxies with stellar masses above $M_\star > 10^{11} \: M_\odot$ undergo on average $0.84^{+0.30}_{-0.20}$ major mergers and $1.43^{+0.50}_{-0.23}$ minor mergers. Moreover, cosmological hydrodynamical simulations (e.g. \citealt{2014Natur.509..177V}) have shown that the average number of mergers for massive galaxies increases with increasing galaxy stellar mass. Thus, the most massive early-type galaxies in the Universe may have experienced up to $\sim5$--$10$ mergers and more with a mass ratio exceeding $10$ since $z\sim3$. The potential lower mass central massive black holes brought in through minor mergers and accretion events that are expected to be more numerous might eventually sink to the center on longer timescales and assemble there without immediate coalescence \citep[see e.g.][]{Partmann_2024}. It is therefore possible that a central SMBH is surrounded by lower mass black holes while a major galaxy merger brings in another SMBH sinking to the center.

Once an additional SMBH enters a galaxy through galaxy merger processes, it can sink towards the remnant center through dynamical friction \citep{Chandrasekhar_1942,Begelman_1980Natur,Tremmel_2015}. Once bound at $\sim10 \mathrm{pc}$, binaries harden via 3-body interactions with stars \citep{Quinlan_1996,Sesana_2006,Sesana_2008a}, a process enhanced in triaxial merger remnants that provide stars with low-angular-momentum orbits capable of interacting with the central SMBH binary \citep{Khan_2016,Gualandris_2017,Bortolas_2018,Rantala2018,Rantala2019}. Such binaries are primary sources of gravitational waves (GWs), which can be individually resolved in the millihertz band by the Laser Interferometer Space Antenna (LISA) \citep{Sesana_2005,Berti_2006,Amaro_Seoane_2013,Amaro_Seoane_2017_LISA,Katz_2020,Mangiagli_2022}, or detected through their collective emission in the form of a gravitational-wave background (GWB) in the nanohertz regime with Pulsar Timing Arrays (PTAs) \citep{Sesana_2008b,Sesana_2009b,Babak_2012,NANOGrav_2025}. Considerable effort has also been devoted to modeling the expected GWB spectrum from massive black hole binary populations using cosmological and semi-analytic simulations, in order to reproduce and interpret PTA measurements \citep[e.g.,][]{Kelley_2017,Fastidio_2025}.

To date, while dozens of SMBH binary candidates have been proposed, only a handful can be considered robust. For example, \cite{Rodriguez_2006} report a resolved binary in the radio galaxy 0402+379 with a projected separation of $\sim7 \,\mathrm{pc}$. \cite{Valtonen_2008} studied OJ287, whose quasi-periodic optical outbursts (12-year cycle, two peaks per cycle) agree to within $\sim1$ day with predictions from a binary SMBH model and are consistent with orbital energy loss via gravitational radiation. In the quasar PG1302-102, \cite{Graham_2015} detected a smooth optical periodicity of $\sim1884$ days ($\sim5.2 \,\mathrm{yr}$) over nearly a decade, which is best explained by a sub-parsec SMBH binary. And more recently \cite{Kharb_2017} reported a candidate dual SMBH at a projected separation of only $\sim0.35 \,\mathrm{pc}$ in the Seyfert galaxy NGC7674 (Mrk533), using VLBI imaging of radio cores. Recently, JWST observation have revealed binary and even triple SMBHs already at redshifts $z\sim 5$ \citep{2024MNRAS.531..355U,2025arXiv250921575U}. However, such studies are not sensitive to SMBH masses lower than $\sim 10^6 M_\odot$. 

Despite the well-established mechanisms of dynamical friction and three-body scattering that drive massive black holes (MBHs) toward the galactic center, these processes do not necessarily guarantee rapid coalescence \citep[e.g.,][]{Partmann_2024}, especially for low-mass MBHs and gas poor environments \citep[e.g.,][]{Van_Wassenhove_2014}. Instead, minor and mini mergers may result into a buildup of a population of MBHs orbiting the central SMBH, which may stall on sub-parsec scales for extended periods of time \citep{Milosavljevic_2001,Dosopoulou_2017}. This accumulation of multiple MBHs in galactic nuclei \citep[e.g.,][]{Matsui_2012,Tanikawa_2014,Askar_2021} can have important implications for both the dynamics of the central region and the rates of subsequent mergers. Motivated by this scenario, we explore the evolution of a system in a galactic nucleus in which a central SMBH is surrounded by a population of MBHs that has formed via minor and mini mergers. The host galaxy of the system may later undergo a major merger that introduces a secondary SMBH into the nucleus, providing an ideal laboratory to study the interplay between hierarchical galaxy assembly and multiple black hole coalescence.

On small spatial scales, the dynamics of stars, stellar mass compact objects (CO) as well as IMBHs in the immediate vicinity (sphere of influence) of an SMBH are governed by a competition between coherent torques, Newtonian mass precession, relativistic precession and stochastic two-body encounters \citep{Rauch_1996, Hopman_2006, Merritt_2011}. In a near-Keplerian potential the orbits of stars can be considered 'wires' acting as long-range gravitating rods that exert slowly varying torques on one another, thus their motion becomes correlated. Such torques lead to the diffusion of orbital angular momentum changing both its direction and magnitude via a process known as \emph{resonant relaxation} (RR; \citealt{Rauch_1996, Hopman_2006, Perets_2009, Merritt_2011, Hamers_2014, Bar-Or_2016}). In dense stellar environments such as galactic nuclei, RR becomes very efficient at randomizing the angular-momentum vectors driving stars and COs to highly eccentric orbits, leading to close passages to the SMBH. When the pericentre distance falls below the tidal radius the star is shredded as a tidal‑disruption event \citep{Rees_1988} or ejected as a hyper-velocity star \citep{Hills_1988}, whereas compact remnants that survive the plunge become extreme- or intermediate- mass‑ratio inspirals (EMRIs/IMRIs) \citep[see e.g.,][for a review]{Amaro_Seoane_2007} inspiralling, into the SMBH hole via gravitational‑wave emission \citep{Merritt_2011, Amaro_Seoane_2018, Broggi_2022, Rom_2024} . In the special case where the inspiral of an IMBH ($10^2 \mathrm{M_{\odot}} \lesssim M_\bullet \lesssim 10^5 \mathrm{M_{\odot}}$) into an SMBH results in a merger, the event is referred to as a \emph{heavy IMRI} \citep{Amaro_Seoane_2023_LISA,Bellovary_2025}.

Building on the analytic picture, a series of numerical studies introduced more realistic treatments of RR including relativistic effects. Using statistical methods such as Fokker–Planck and Monte Carlo simulations, \citet{Merritt_2011} demonstrated that the inclusion of post-Newtonian (PN) corrections substantially alters the dynamical evolution of compact objects around SMBHs. In particular, the relativistic precession induced by the PN terms suppresses the efficiency of resonant torques in driving angular momentum diffusion. As a consequence, the rate at which objects can be driven to the EMRI regime is significantly reduced. This dynamical suppression was termed the \emph{Schwarzschild barrier}, a phenomenon that has since been independently confirmed and further explored in subsequent studies \citep{Brem_2014,Hamers_2014}. 

EMRIs constitute one of the key science objectives of space-based gravitational-wave observatories such as LISA \citep{Babak_2017, Amaro_Seoane_2017_LISA}. Their gradual orbital evolution and the accumulation of a very large number of cycles around an MBH allow for exceptionally high signal-to-noise ratios (SNR) \citep{Amaro_Seoane_2018}. This, in turn, enables precise mapping of the spacetime geometry in the strong-gravity regime \citep{Ryan_1995}, making EMRIs powerful probes for testing the predictions of General Relativity (GR) \citep{Pappas_2015, Cardoso_2016, Cardenas_2024}. Furthermore, since their orbital evolution and consequently their GW-waveform is extremely sensitive to external perturbations, such events can also provide valuable information on: the presence of a an MBH companion \citep{Yunes_2011a} or a gaseous disc in the system \citep{Narayan_2000,Yunes_2011b}.

An important quantity in studies of compact object dynamics around massive black holes is the relative occurrence of direct plunges (DPs) versus EMRIs. In the classical picture, DPs are expected to arise mainly from orbits with large initial semi-major axes, while EMRIs originate from more tightly bound orbits with smaller semi-major axes. Because DPs follow highly eccentric, nearly radial trajectories, they emit only weak gravitational radiation, making their signals difficult to detect with current GW detectors due to the resulting low signal-to-noise ratio (SNR). However, \citet{Qunbar_2024} recently showed that this dichotomy breaks down for lower-mass SMBHs ($M_{\bullet} \lesssim 10^5 \: \mathrm{M_{\odot}}$). In this regime, they identified a new class of inspirals—termed \emph{cliffhanger} EMRIs—that do not follow the standard semi-major axis dependence, subsequently confirmed by \cite{Mancieri_2025a}. More broadly, recent work has highlighted a growing diversity of EMRI formation channels and phenomenology, including highly eccentric burst-like inspirals with extremely large mass ratios (X-MRIs; \citealt{Amaro_Seoane_2019}), environmentally driven EMRIs with long inspiral timescales (E-EMRIs; \citealt{Amaro_Seoane_2024}), and stochastic, intermittently observable sources often referred to as \emph{popcorn} EMRIs \citep{Amaro_Seoane_2025_popcorn}, underscoring that the traditional DP–EMRI classification may be overly simplistic.

In analogy to the dynamical evolution around a single SMBH, the presence of a second SMBH introduces additional layers of complexity, giving rise to a broader range of rich and distinctive dynamical phenomena. In such systems, the interplay between two-body relaxation, relativistic precession, and secular mechanisms such as the eccentric von Zeipel-Kozai–Lidov (ZKL) oscillations \citep{von_Zeipel_1910,Kozai_1962, Lidov_1962, Naoz_2016,Ito_2019}, combined with the inherently chaotic nature of multi-body interactions, can significantly alter the orbital evolution of surrounding compact objects and stars. It has been found that this channel can enhance the rate of EMRIs \citep{Bode_2013, Naoz_2022, Mazzolari_2022} and tidal disruption events (TDEs) \citep{Chen_2008,Chen_2009,Chen_2011,Chen_2013,Li_2015,Ricarte_2016}, thereby enriching the astrophysical signatures expected from galactic nuclei hosting SMBH binaries.

Building on these ideas, \citet{Bode_2013, Naoz_2022, Mazzolari_2022} examined scenarios in which a primary SMBH, comparable in mass to Sgr A$^*$ at the Galactic Centre \citep{GRAVITY_2020, EHT_2022}, is embedded within a nuclear star cluster (NSC) of stars and compact objects, while the mass of a secondary MBH is varied. Using semi-analytic methods and direct integrations of the secular hierarchical three-body equations of motion, these studies demonstrated that the presence of a second MBH can substantially increase the rate of EMRI formation relative to the single-SMBH case, with the ZKL mechanism acting as the dominant driver. More recently, \citet{Naoz_2023} extended this line of investigation to assess observational prospects, concluding that LISA could potentially detect several hundred such EMRIs with SNR$\geq 8$ over a four-year mission.

Our study departs from previous works on EMRI formation and the dynamics around SMBHs in galactic nuclei in several key aspects: (i) we consider systems hosting a very massive SMBH of $M_\bullet = 10^9 \: \mathrm{M_{\odot}}$ surrounded by a small cluster of $10^5 \: \mathrm{M_{\odot}}$ MBHs, formed as the outcome of successive minor/mini galaxy mergers and accretion events; (ii) the secondary, equal-mass SMBH is introduced through a major galaxy merger; and (iii) the dynamical evolution of these black hole systems is followed using direct $N$-body simulations that incorporate relativistic corrections up to 3.5PN order. A setup similar to our single-SMBH runs, but involving lower-mass SMBHs and MBHs, was recently examined by \cite{Hochart_2024}, who focused on how the PN formalism accelerates the ejection of MBHs and how varying SMBH masses can favor ejections over mergers. In this paper, we investigate merger rates in the presence of single versus binary SMBHs and examine the differences in their formation pathways, focusing on the relative fractions of DPs and IMRIs, as well as their detectability with LISA. We find that the presence of a second SMBH leads to a substantial enhancement in merger events, by a factor of $2$–$5$, with the majority corresponding to highly eccentric DPs. We also find that, although the absolute number of IMRIs is lower in the single-SMBH runs, their relative fraction is larger compared to the binary-SMBH case.

The paper is organized as follows. In Section~\ref{sec:timescales}, we provide an overview of the theoretical framework for orbital evolution and dynamical timescales around SMBHs. Section~\ref{sec:methods} presents our $N$-body code, including a new adaptive timestep implementation and two complementary definitions of orbital elements suited for resolving both DPs and IMRIs. In Section~\ref{sec:dynamical_evol}, we analyze the evolution of our systems across different dynamical regimes and discuss the resulting merger demographics and classifications. Section~\ref{sec:GWs} explores the gravitational-wave frequency bands of our mergers and their detectability prospects. Finally, Section~\ref{sec:Conclusions} summarizes our conclusions and outlines directions for future work.

\section{Theoretical Background - Motion around Supermassive Black Holes}
\label{sec:timescales}

\begin{table}
\centering
\renewcommand{\arraystretch}{1.1} 
\begin{tabular}{l c c c}
\hline
Physical mechanism & Timescale & $N=1$ [yr] & $N=10$ [yr]\\
\hline
Keplerian period & $P$ & $3.0$ & $3.0$ \\

Mass precession & $t_{\mathrm{MP}}$ & $1.71 \times 10^4$ & $1.71 \times 10^3$ \\
Schwarzschild precession & $t_{\mathrm{GR}}$ & $77.4$ & $77.4$ \\
Resonant relaxation & $t_{\mathrm{RR}}$ & $4.71 \times 10^3$ & $1.49 \times 10^3$ \\
Non-resonant relaxation & $t_{\mathrm{NR}}$ & $2.96 \times 10^8$ & $2.96 \times 10^7$ \\
GW inspiral & $t_{\mathrm{GW}}$ & $8.75 \times 10^7$ & $8.75 \times 10^7$ \\
von Zeipel-Kozai-Lidov & $t_{\mathrm{ZKL}}$ & $3.27 \times 10^5$ & $3.27 \times 10^5$ \\
\hline
\end{tabular}
\caption{Characteristic timescales of different dynamical processes for an example system: $N=10$ lower-mass MBHs around the primary SMBH of mass $M_\bullet$. Primary and secondary mass $M_{\bullet}=10^9 \mathrm{M_{\odot}}$, $m=10^5 \mathrm{M_{\odot}}$, semi-major axis $a=0.01 \mathrm{pc}$, eccentricity $e=0.5$ for two different values of the number of MBHs $N$ inside $a$. For $t_{\mathrm{ZKL}}$, we consider an equal mass companion SMBH, with on an outer orbit at $a_{\mathrm{out}}=1 \mathrm{pc}$ and $e=0.5$.}
\label{tab:timescale}
\end{table}

To better understand the motion and dynamics of a system of $N$ bodies in the vicinity of a single SMBH, in this section we brifely describe the basic concepts of resonant relaxation \citep{Rauch_1996, Hopman_2006}, precession due to general relativity \citep{Einstein_1915, Weinberg_1972} and the eccentric Kozai-Lidov effect \citep{von_Zeipel_1910, Kozai_1962,Lidov_1962,Naoz_2016} induced by a tertiary object. Additionally we outline the characteristic timescales of each process, which are crucial for identifying the dominant mechanisms under different conditions.

\subsection{Mass Precession} 

For a purely Keplerian orbit the radial $P_\mathrm{r}$ and angular $P_\mathrm{\phi}$ orbital periods correspond to the same motion, so $P_\mathrm{r}=P_\mathrm{\phi}$. In the presence of a distribution of mass around the SMBH those two periods are not the same anymore, leading to retrograde precession of the argument of periastron $\omega$. For a spherically symmetric mass distribution the time $\omega$ takes to precess by $\pi$ is approximately \citep{Rauch_1996,Merritt_2013_book}

\begin{equation}
\label{eq:tMP}
    t_{\mathrm{MP}} \approx \frac{1}{2}(1-e^2)^{-1/2}\frac{M_\bullet}{M(a)}P,
\end{equation}
where $M(a)$ is the distributed mass inside $a$ and therefore the process is called \emph{mass precession} (MP). For a system of $N = 10$ MBHs of mass $m = 10^5 \: \mathrm{M_{\odot}}$ orbiting a primary SMBH of mass $M = 10^9 \: \mathrm{M_{\odot}}$, an MBH with semi-major axis $a = 0.01  \mathrm{pc}$ and eccentricity $e = 0.5$ would require $t_{\mathrm{MP}} = 1710 \: \mathrm{yr}$ for MP to significantly affect its orbit (see Table~\ref{tab:timescale}).

\subsection{Relativistic Precession} 
For a non-spinning SMBH, the leading-order effect of general relativity to the orbital motion leads to prograde precession of $\omega$. This so called \emph{Schwarzschild precession} (or GR-precession) \citep{Einstein_1915} does not affect the orientation of the orbital plane, while the time needed for $\omega$ to precess by $\pi$ is \citep{Weinberg_1972,Rauch_1996,Merritt_2013_book}

\begin{equation}
\label{eq:tGR}
t_{\mathrm{GR}} = \frac{1}{6}(1-e^2)\frac{c^2a}{GM_\bullet}P,
\end{equation}

where $c$ is the speed of light. In a system of $N = 10$ MBHs with individual masses $m = 10^5 \: \mathrm{M_{\odot}}$ orbiting a primary SMBH of mass $M = 10^9 \:  \mathrm{M_{\odot}}$, an MBH with semi-major axis $a = 0.01  \mathrm{pc}$ and eccentricity $e = 0.5$ would experience relativistic precession on a timescale of $t_{\mathrm{GR}} = 7.74 \: \mathrm{yr}$ (Table~\ref{tab:timescale}), making GR precession by far the most rapid and dominant process in such a configuration.

\subsection{Resonant Relaxation}
In galactic nuclei dominated by a central point mass $M_\bullet$ (e.g. an SMBH), orbits are nearly Keplerian, and their orientations remain fixed over many orbital periods. This leads to the phenomenon of \emph{resonant relaxation} (RR) \citep{Rauch_1996,Hopman_2006,Perets_2009,Merritt_2011,Hamers_2014,Bar-Or_2016}, whereby small, long-lived torques between objects result in a rapid evolution of angular momentum vectors. In the \emph{coherent} RR regime, these torques act in a fixed direction over a coherence time $t_{\mathrm{coh}}$ (the time for which orbits stay fixed), leading to a net change in angular momentum that grows linearly with time. The total torque $T$ on a test particle due to $N$ randomly oriented background ones at radius $r\lesssim a$ is of order
\begin{equation}
T \sim \frac{Gm\sqrt{N}}{a},
\end{equation}
where $G$ is Newton's gravitational constant, $m$ is the average mass of background objects and $a$ the semi-major axis of the test particle. On timescales shorter than $t_{\mathrm{coh}}$, this leads to an approximately linear angular momentum change of the order of $|\dot{\textbf{L}}| \approx \sqrt{N}(Gm/a)$, while $|\dot{E}| \approx 0$.  Expressed in terms of orbital period $P$ and the angular momentum of a circular orbit $L_\mathrm{c}=\sqrt{GM_\bullet a}$ the change in $L$ for $\Delta t<t_{\mathrm{coh}}$ is then \citep{Merritt_2013_book},

\begin{equation}
    \frac{\Delta L}{L_{\mathrm{c}}} \approx\sqrt{N}\frac{GM}{a} \times \frac{\Delta t}{\sqrt{GM_\bullet a}} \approx 2\pi \frac{m\sqrt{N}}{M_\bullet}\frac{\Delta t}{P}.
\end{equation}

The change of $\textbf{L}$ translates to the precession of the orbital plane, on a characteristic timescale \citep{Rauch_1996}

\begin{equation}
\label{eq:tRR}
    t_{\mathrm{RR}} \equiv \frac{P}{2\pi}\frac{M_\bullet}{m}\frac{1}{\sqrt{N}}.
\end{equation}

In a system of $N = 10$ MBHs with individual masses $m = 10^5 \:\mathrm{M_{\odot}}$ orbiting a primary SMBH of mass $M = 10^9 \:\mathrm{M_{\odot}}$, an MBH with semi-major axis $a = 0.01 \: \mathrm{pc}$ and eccentricity $e = 0.5$ would experience relativistic precession on a timescale of $t_{\mathrm{RR}} = 1490 \: \mathrm{yr}$ (Table~\ref{tab:timescale}), competing with $t_{\mathrm{MP}}$ while both mechanisms being suppressed by GR. 

\subsection{Non-resonant (two-body) relaxation} 
In contrast to resonant relaxation, classical two-body (non-resonant) relaxation arises from uncorrelated, close gravitational encounters. It affects both the energy and angular momentum of stars, and its timescale is much longer than that of resonant processes in a Keplerian potential. This leads to the change of orbital elements on a timescale given by \citep{Rauch_1996,Hopman_2006}

\begin{equation}
\label{eq:tNR}
t_{\mathrm{NR}} \approx \frac{M_\bullet^2}{m^2 N} P,
\end{equation}

and is typically orders of magnitude larger than both $t_{\mathrm{MP}}$ and the timescale of relativistic precession $t_{\mathrm{GR}}$ which we discuss in next sub-section. Although $t_{\mathrm{NR}}$ is orders of magnitude (see Table~\ref{tab:timescale}) longer than both $t_{\mathrm{MP}}$ and $t_{\mathrm{GR}}$, it is important to note that, unlike resonant relaxation, non-resonant relaxation is not suppressed by relativistic precession. This is because it arises from stochastic, uncorrelated two-body encounters rather than coherent torques, and thus remains effective even in regimes where GR rapidly quenches resonant processes \citep{Rauch_1996,Hopman_2006}.

\subsection{Gravitational Wave Inspiral}
Compact binaries emit gravitational waves, which remove energy and angular momentum from the system, causing the orbit to shrink and circularize over time. Once a particle is sufficiently close to the SMBH and decoupled from the rest of the system, the characteristic timescale for the semi-major axis to decay due to GW emission is given by \cite{Peters1963} as

\begin{equation}
\label{eq:tGW}
t_{\mathrm{GW}} = \frac{5}{64} \frac{c^5 a^4}{G^3 M m (M+m)} F(e),
\end{equation}

where 

\begin{equation}
\label{eq:F(e)}
    F(e)=(1-e^2)^{7/2} \left(1 + \frac{73}{24} e^2 + \frac{37}{96} e^4 \right)^{-1}
\end{equation}

and $a$ and $e$ are the orbital semi-major axis and eccentricity, and $m_1$, $m_2$ are the component masses. This timescale sets the orbital decay rate for GW-driven inspiral and is strongly dependent on eccentricity. 

Starting from the post-Newtonian equations of motion of a binary system, it is possible to (numerically) solve the GW merger timescale of the system. For highly eccentric and relativistic orbits the GW timescale defined above begins to deviate from the merger timescale \citep{Vazquez_Aceves_2021}. A more general expression, valid for arbitrary eccentricities, was derived by \citet{Peters1964}, with an alternative formulation later presented by \citet{Bonetti_2018}. More recently, \citet{Zwick_2020} introduced a correction factor $R(e)$ ,

\begin{equation}
\label{eq:R(e)}
    R\left(e\right)=8^{1-\sqrt{1-e}},
\end{equation}

which, when applied to Eq.~\eqref{eq:tGW}, yields a timescale applicable across the full eccentricity range:

\begin{equation}
\label{eq:tGW_Zwick}
t_{\mathrm{GW}}^{\mathrm{corr}} = t_{\mathrm{GW}} R(e).
\end{equation}

As an example, for a system with primary and secondary mass $M_{\bullet}=10^9 \mathrm{M_{\odot}}$, $m=10^5 \mathrm{M_{\odot}}$, semi-major axis $a=0.01 \mathrm{pc}$, eccentricity $e=0.5$, $t_{\mathrm{GW}}^{\mathrm{corr}}=87.5 \: \mathrm{Myr}$ , while  $t_{\mathrm{GW}}=47.6 \: \mathrm{Myr}$. In the following we adopt this definition for the GW-driven inspiral timescale and $t_{\mathrm{GW}}$ will therefore correspond to Eq.\eqref{eq:tGW_Zwick}. 

\subsection{Eccentric von Zeipel-Kozai-Lidov Mechanism}
Let us now consider the scenario where a second SMBH with mass $M_{2}$ at a sufficiently large distance forms a binary with the primary SMBH $M_{1}$ and semi-major axis of $a_{\mathrm{out}}$. Now assume that a test particle (in our case a lower-mass MBH) with mass $m$ is orbiting the SMBH with semi-major axis $a_{\mathrm{in}}$, then a hierarchical three-body system is formed. A hierarchical triplet consists of two clearly separated binary orbits, i.e., $a_{\mathrm{in}}/a_{\mathrm{out}} \ll 1$ , where $a_{\mathrm{in}}$ is the semi-major axis of the inner binary. In this scenario the outer SMBH acts as a perturber leading to the exchange of angular momentum between the two orbits \citep{von_Zeipel_1910,Kozai_1962,Lidov_1962}. Especially, when the relative inclination between the two orbital planes is above a critical threshold $(i_{\mathrm{crit}}>39.2^{\circ})$ , oscillations between the inner binary eccentricity and the relative inclination are triggered, acting on a (quadrupole approximation) timescale \citep{Antonini_2015} 

\begin{equation}
\label{eq:tZKL}
t_{\mathrm{ZKL}} \sim \frac{16}{30 \pi} \frac{M_{1}+M_{2}+m}{M_{2}} \frac{P_{\mathrm{out}}^2}{P_\mathrm{in}}\left(1-e_{\mathrm{out}}^2\right)^{3 / 2},
\end{equation}

 where $e_{\mathrm{out}}$ is the outer binary eccentricity and $P_\mathrm{in}$ and $P_\mathrm{out}$ the inner and outer orbital periods. The ZKL mechanism can induce large eccentricity oscillations on a timescale longer than $P_\mathrm{2}$, potentially driving the inner orbit to very small pericenter distances. This enhances the likelihood of direct collisions \citep{Thompson_2011,Hamers_2013,Antonini_2017,Silsbee_2017,Toonen_2018,Hamers_2019,Liu_2019} or rapid gravitational-wave–driven mergers, provided that GR precession does not suppress the oscillations \citep{Holman_1997,Bonetti_2016,Lim_2020}. Hierarchical triples that become dynamically unstable ($r_\mathrm{p,out}/a_\mathrm{in} \sim 1$) can also trigger collisions, mergers, or ejections. As an example, for an SMBH binary on an outer eccentric orbit with $a_{\mathrm{out}} = 1\:\mathrm{pc}$, $e_{\mathrm{out}} = 0.5$, equal masses $M_1 = M_2 = 10^9 \:\mathrm{M_{\odot}}$, and an inner MBH of $m = 10^5 \: \mathrm{M_{\odot}}$, the timescale for the tertiary to induce ZKL oscillations on the inner orbit is $t_{\mathrm{ZKL}} \sim 3.27 \times 10^4 \: \mathrm{yr}$.

\section{Methods}
\label{sec:methods}

We explore the post-Newtonian dynamics and mergers of a system of massive black holes ($m = 10^5 M_\odot$) around a supermassive black hole ($M_\bullet = 10^9 M_\odot$). In half of our models a second SMBH is introduced on an outer orbit around the primary SMBH and its surrounding MBH system. To investigate the role of the spatial distribution of 10 MBHs around the primary SMBH and the impact of a companion SMBH on the merger demographics we have performed a series of 300 direct $N$-body simulations. The systems are evolved for 10 $\mathrm{Myr}$ with the use of the regularised post-Newtonian \mstar{} integrator \citep{Rantala2020} included in the direct $N$-body code \bifrost{} \citep{Rantala2023}. \bifrost{} is used for handling the input, on-the-fly analysis and outputs of the simulations but we do not use its main fourth order forward integrator in this study. In the following section, we briefly outline the main features of \mstar{} and highlight the modifications we made to accurately capture plunging and highly relativistic inspiraling orbits.

\subsection{$N$-body Code}
\label{section:n-body}
Throughout this study we utilized the \mstar{} integrator \citep{Rantala2020} included in the direct-summation $N$-body code \bifrost{} \citep{Rantala2023}. \mstar{} enables the accurate modelling non-softened gravitational dynamics around SMBHs using algorithmically regularised integration \citep{Mikkola1999,Mikkola_2006,Mikkola2008, Rantala2017}. The inclusion of relativistic corrections in direct $N$-body simulations has a long history, with pioneering work by \citet{Kupi_2006} introducing for the first time post-Newtonian (PN) terms into $N$-body integrators for compact-object dynamics, later extended to include spin-related effects by \citet{Brem_2013}.

The algorithmic regularisation relies on three key numerical aspects. First, the equations of motion of the few-body systems are time transformed. Together with a leapfrog approach it is ensured that the Newtonian coordinate singularity at small separations is avoided. Next, the particles are organized in a minimum spanning tree coordinate structure \citep{Rantala2020} which further reduces floating-point round-off errors \citep{Mikkola_1993,Mikkola1999}. The final aspect of the algorithm is the use of Gragg-Bulirsch-Stoer (GBS) \citep{Gragg_1965,Bulirsch1966} extrapolation method. The GBS method ensures the high accuracy of numerical orbit integrations with the relative errors in the dynamical variables of the system being below a user-defined error tolerance $(\eta_{\mathrm{GBS}})$ .

 \subsection{Relativistic dynamics at 3.5PN order}

The post-Newtonian formalism is a perturbative expansion technique applied to Einstein’s field equations in the regime of weak gravitational fields and low velocities. In this approach, relativistic effects —originally encoded in the geodesic equations— are systematically introduced as higher-order corrections to Newtonian gravity. For a system characterized by typical velocity $v$ , mass $M$ and length scale $R$, the expansion is governed by a small parameter $\epsilon$ indicative of the deviation from the Newtonian limit 

\begin{equation}
\epsilon \sim\left(\frac{v}{c}\right)^2 \sim\left(\frac{G M}{c^2 R}\right),
\end{equation}

where $G$ is Newton's gravitational constant and $c$ is the speed of light. For $\epsilon \rightarrow 0$ we recover the Newtonian gravity, while $\epsilon \sim 1$ corresponds to the strong gravitational regime.  The post-Newtonian correction terms of the order iPN are denoted such that each term scales with a specific power of the PN expansion parameter $\epsilon$,

\begin{equation}
\left|\boldsymbol{a}_{i \mathrm{PN}}\right| \propto \epsilon^i \sim\left(\frac{v}{c}\right)^{2 i} \sim\left(\frac{R_\mathrm{sch}}{R}\right)^i,
\end{equation}

where $v$ and $R$ are now the relative velocity and separation between two BH particles and $R_{\mathrm{sch}}=2GM/c^2$ is the Schwarzschild radius for a binary with total mass $M$. The PN-corrected acceleration can then be written as

\begin{equation}
\boldsymbol{a}=\boldsymbol{a}_{\mathrm{N}}+\boldsymbol{a}_{\mathrm{PN} 1}+\boldsymbol{a}_{\mathrm{PN} 2}+\boldsymbol{a}_{\mathrm{PN} 3}+\boldsymbol{a}_{\mathrm{PN} 2.5}+\boldsymbol{a}_{\mathrm{PN} 3.5},
\end{equation}

where $\boldsymbol{a}_{\mathrm{N}}$ is the Newtonian acceleration. Terms associated with conservation of energy and angular momentum are called \textit{conservative} and represented by integer order, while \textit{dissipative} terms with half-integer terms are related to radiation reaction due to gravitation wave emission.

To account for general-relativistic effects of the orbital motion around SMBHs \citep[e.g.][]{Will_2006,Poisson_2014}, \mstar{} includes an implementation of the post-Newtonian velocity-dependent equations of motion \citep{Rantala2017,Mannerkoski2021,Mannerkoski2023}, following the pioneering work by \cite{Kupi_2006} who introduced PN terms into direct-summation $N$-body codes. The velocity-dependent equations of motion are solved using an auxiliary velocity algorithm \citep{Hellstrom_2010} which allows for the efficient implementation of the PN-corrected gravitational dynamics. We incorporate post-Newtonian correction terms up to 3.5PN order, derived for binary systems with arbitrary eccentricity in the modified harmonic gauge, following the formulation presented in \citep{Mora_2004}. Furthermore, we include the relevant cross terms, which are essential for binary systems that are not isolated, particularly when the timescales of encounters or hierarchical triples are comparable to the relativistic pericenter-advance timescale \citep{Will_2014}, or when they induce a cumulative gravitational-wave phase shift of several $2\pi$ over timescales shorter than one year, as in EMRI/IMRI systems \citep{Amaro_Seoane_2025}. Finally, for the purposes of this study, we neglect black hole spin–dependent terms.

\subsection{Post-Newtonian Orbital Elements}
In Newtonian gravity, the orbit of a binary system can be described using Keplerian orbital elements such as the semi-major axis and eccentricity, which remain constant or vary only gradually due to external perturbations. However, when PN corrections are included —as is necessary for the orbital evolution of BH binaries— the standard Keplerian elements are no longer even approximately conserved over a single orbit \citep{Damour_1985}. Instead, they exhibit significant oscillations, particularly near pericenter in eccentric orbits \citep{Damour_1985,Danby_1992}. At short binary separations where the orbit becomes highly relativistic, the Keplerian elements no longer describe the geometry of the orbit. This mismatch leads to a commonly found artifact of increasing Keplerian eccentricity towards merger \citep[see e.g.,][]{Mannerkoski_2019}. To overcome this, one has to introduce some quasi-Keplerian parametrization which takes into account the PN-corrected orbital motion. We have implemented such parametrization in \mstar{}, by computing on-the-fly the 3PN-accurate quasi-Keplerian elements following \cite{Memmesheimer_2004}. In previous studies using \mstar, the instantaneous post-Newtonian orbital elements have been computed in post processing from the code output (e.g. \citealt{Mannerkoski_2019, Mannerkoski2021}). The quasi-Keplerian elements correspond to the solution of the conservative part of the PN equations of motion for a non-spinning binary:

\begin{equation}
R=a\left(1-e_R \cos u\right),
\end{equation}

\begin{equation}
\begin{aligned}
n\left(t-t_0\right)= & u-e_t \sin u+\left(\frac{g_{4 t}}{c^4}+\frac{g_{6 t}}{c^6}\right)(v-u) \\
& +\left(\frac{f_{4 t}}{c^4}+\frac{f_{6 t}}{c^6}\right) \sin v+\frac{i_{6 t}}{c^6} \sin 2 v \\
& +\frac{h_{6 t}}{c^6} \sin 3 v,
\end{aligned}
\end{equation}

\begin{equation}
\begin{aligned}
\frac{2 \pi}{\Phi}\left(\phi-\phi_0\right)= & v+\left(\frac{f_{4 \phi}}{c^4}+\frac{f_{6 \phi}}{c^6}\right) \sin 2 v \\
& +\left(\frac{g_{4 \phi}}{c^4}+\frac{g_{6 \phi}}{c^6}\right) \sin 3 v+\frac{i_{6 \phi}}{c^6} \sin 4 v \\
& +\frac{h_{6 \phi}}{c^6} \sin 5 v,
\end{aligned}
\end{equation}

\begin{equation}
\tan \frac{v}{2}=\sqrt{\frac{1+e_\phi}{1-e_\phi}} \tan \frac{u}{2}.
\end{equation}

The coordinates $R$ and $\phi$ represent the relative separation and angle in the orbital plane, with $\phi_0$ setting the pericenter direction at $t = t_0$. The orbit remains elliptical with semi-major axis $a$, but the radial frequency $f_r = n / 2\pi$ differs from the angular frequency, causing the angle $\phi$ to precess by $\Phi > 2\pi$ per orbit. The anomalies $u$ and $v$ generalize the eccentric and true anomaly, but involve additional PN correction factors ($g$, $f$, $i$, $h$). The Newtonian eccentricity $e$ is replaced by three distinct ones: $e_R$, $e_t$, and $e_\phi$, i.e., eccentricity is not a scalar anymore. In the Newtonian limit where PN corrections are negligible, this reduces to the standard Keplerian description.
The conserved energy $E$ and angular momentum $L$ at 3PN (conservative) order are given by \cite{Memmesheimer_2004},

\begin{equation}
\label{eq:EL_PN3}
\begin{gathered}
E=\mu\left(E_0+c^{-2} E_1+c^{-4} E_2+c^{-6} E_3\right) \\
L=\mu|\boldsymbol{R} \times \boldsymbol{V}|\left(L_0+c^{-2} L_1+c^{-4} L_2+c^{-6} L_3\right),
\end{gathered}
\end{equation}

where $\mu = m_1m_2/M$ is the reduced mass and $M=m_1+m_2$ the total mass of the binary and $\boldsymbol{R}$ and $\boldsymbol{V}$ the relative separation and velocity vectors computed in \mstar{}. Even in the absence of radiation reaction terms, $E$ and $L$ are not constant in time, leading to the oscillation of the corresponding 3PN orbital elements. Additional oscillatory behavior is induced when radiation reaction terms are included \citep{Koeningsdoerffer_2006}, but we expect those to be negligible compared to the secular evolution. 
The components of the eccentricity vector do not exactly relate to the geometry of a Keplerian orbit, which makes their interpretation ambiguous. The time eccentricity $e_t$ remains well defined, but it exhibits a modest growth as the system approaches merger. This increase, however, does not reflect a genuine rise in the physical eccentricity of the orbit, i.e. the orbit does not become more elongated when closing the merger. For nearly circular relativistic orbits, the 3PN expressions yield $e_R^{2}<0$ and $e_{\phi}^{2}<0$ \citep{Mannerkoski_2019}, while $e_t$ shows a behavior reminiscent of the Keplerian case without corresponding to a real eccentricity growth. For further discussion, see \cite{Mannerkoski_2019}.

\subsubsection{Geometric Eccentricity}
\begin{figure}
	\includegraphics[width=0.9\columnwidth]{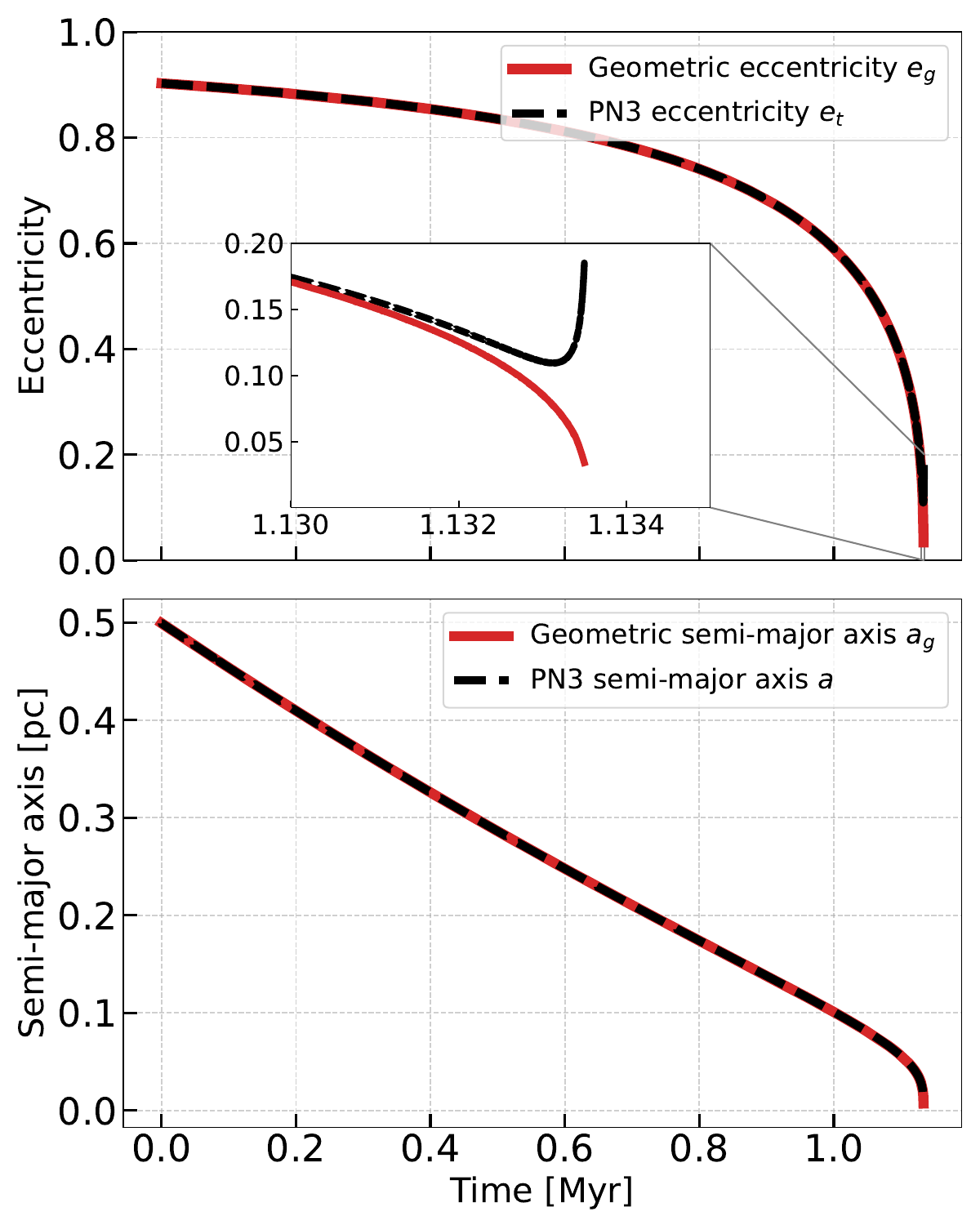}
    \caption{Orbital evolution of a merging SMBH binary with $M_1=10^{10} \: \mathrm{M_{\odot}}$, $M_2=10^{9} \: \mathrm{M_{\odot}}$. The initial eccentricity and semi-major axis are $e=0.9.$ and $a=0.5 \: \mathrm{pc}$ respectively. The inset panel in the top panel highlights the well behaved geometric eccentricity close to merger. The increase of the 3PN radial component $e_R$ does not correspond to a rise of the orbital eccentricity in the geometric sense.}
    \label{fig:pn3geom}
\end{figure}

A more robust and gauge independent way to measure orbital eccentricity, which also applies to the strong relativistic regime can be defined by utilizing the pericentre $R_{\mathrm{min}}$ and apocentre $R_{\mathrm{max}}$ of a binary orbit \citep{Csizmadia_2012}, given by

\begin{equation}
\label{eq:e_g}
e_g=\frac{R_{\max }-R_{\min }}{R_{\max }+R_{\min }},
\end{equation}

and the corresponding semi-major axis $a_g$ is then
\begin{equation}
a_g=\frac{R_{\max }+R_{\min }}{2}.
\end{equation}

This definition naturally accounts for the eccentricity evolution of binaries, while also encompassing the effects of precession and spin \citep{Phukon_2025} and it has recently been used in the late-time inspiral of binary BHs in general-relativistic-magneto-hydrodynamical (GR-MHD) simulations \citep{Ressler_2025}.
Notice that adopting the above definition, the orbital elements cannot be anymore computed on-the-fly using the instantaneous relative position and velocity vectors of a binary during a simulation run. Rather, a timeseries of the positions over at least a single orbital period is required. It is essential to ensure that the orbit output is sampled with sufficiently high time resolution so that the measurements of $R_{\mathrm{max}}$ and especially $R_{\mathrm{min}}$ are reliable. This is particularly important for strongly precessing orbits, where accurately capturing $R_{\mathrm{min}}$ may require adaptively reducing the output timestep as the system approaches merger. In our simulations we have adopted the geometrical orbital elements description for the binary BH merging orbits and in the next section we discuss our approach and criteria for the adaptive output timesteps. In Fig.~\ref{fig:pn3geom} we provide an example of an eccentric binary SMBH merger initiated with eccentricity $e=0.9.$, semi-major axis $a=0.5 \: \mathrm{pc}$ and component masses $M_1=10^{10} \: \mathrm{M_{\odot}}$, $M_2=10^{9} \: \mathrm{M_{\odot}}$. The ill-behaved rise of the 3PN eccentricity $e_t$ (red curve) close to merger is highlighted in the inset plot of Fig.~\ref{fig:pn3geom}. This distinction is important, since spurious numerical eccentricity near merger could be misinterpreted as evidence for genuinely eccentric gravitational-wave events. Correctly identifying true eccentric mergers is crucial, as they provide valuable insights into the astrophysical environments and dynamical histories of the merging binaries.

\subsection{Adaptive Output Timestep}
\label{sec:adaptive_timestep}
The GBS extrapolation method in \mstar{} is responsible for the adaptive timestepping and order control of the integration \citep{Rantala2020,Mannerkoski2023}. Compared to non-regularised techniques, regularised integrators are highly efficient and only require $\lesssim10$ steps per orbit (especially for non-relativistic binaries) while still retaining a high integration accuracy. However, for the purposes of determining the geometric eccentricity this sparse stepping is not optimal. In our simulations, the final MBH orbits around the primary SMBH can enter the highly relativistic regime, where the precession is very rapid. Our goal is to ultimately distinguish between plunging or inspiral orbits, where the evolution of orbital elements close to merger has to be properly captured \citep{Hochart_2024}. At the same time, we wish to make sure that orbits with large separations are still handled by the default integration timesteps of \mstar{} to achieve optimal performance, i.e., to avoid unnecessary slow-downs of the code when high-frequency output is not needed. For this, we introduce an adaptive time-stepping scheme that shortens the standard \mstar{} integration step length if a better sampled orbit output is required. To do so, we start by finding the minimum relative distance $d_{\mathrm{min}}$ between MBH and SMBH particles and compute their combined Schwarzschild radius

\begin{equation}
\label{eq:Rsch_comb}
    R_{\mathrm{sch}} = \frac{2G (M_\bullet + m)}{c^2},
\end{equation}

where $M_\bullet$ and $m$ are the masses of the SMBH and the MBH respectively. To test whether the original timestep $dt_{0}$ (labelled $H$ in \citealt{Rantala2020}) is capable of capturing the shortest-period close orbit, we compute the orbital $P$ period of a circular orbit at $d_{\mathrm{min}}$,

\begin{equation}
    P(r=d_{\mathrm{min}}) = \frac{2 \pi d_{min}^3}{G (M_\bullet + m)}.
\end{equation}

We can now directly compare the orbital period $P$ to the base timestep $dt_{\mathrm{0}}$ and determine whether a shorter timestep is required for resolving the geometry of the orbit from the code output. As a rule of thumb, we aim to resolve each orbit with at least $N_{\mathrm{points}} = 20$ outputs. However, to better adapt the output time resolution to the dynamics near the SMBH, we define three refinement zones based on the minimum relative distance to the binary, measured in units of the Schwarzschild radius $R_{\mathrm{S}}$:

\[
N_{\mathrm{points}}(r) =
\begin{cases}
100, & r < 25\,R_\mathrm{sch} \\
50,  & 25\,R_\mathrm{sch} \leq r < 52\,R_\mathrm{sch} \\
20,  & 52\,R_\mathrm{sch} \leq r < 236\,R_\mathrm{sch}.\\
\end{cases}
\]

These values serve as target output time resolutions: if an MBH is orbiting the SMBH at short distance, the orbit has to be increasingly better resolved due to rapid precession near the SMBH. For context, a binary SMBH-MBH orbit with $M_\bullet= 10^9 \: \mathrm{M_{\odot}}$ and $m= 10^5 \: \mathrm{M_{\odot}}$ at $r = 236 \, R_\mathrm{sch}$ typically have periods $P \gtrsim 10$ yr, while at $r = 52 \, R_\mathrm{S}$ we have $P \gtrsim 1$ yr. The innermost zone corresponds to the final inspiral stage, where maximal time resolution is critical. We have tested the adaptive output timestep criterion in practice and found that it performs well for all merger orbits in our simulation sample including eccentric mergers.

In the code we next check if $ P(r=d_{\mathrm{min}}) \leq 10 \: dt_{\mathrm{0}} $ and if so, a refined timestep is adopted. The new timestep is estimated as

\begin{equation}
    dt_{\mathrm{new}} = \frac{P}{N_{\mathrm{points}}}.
\end{equation}

This ensures that the effective number of points per orbit is at least $N_{\mathrm{points}}$. 
We have also included the option to cap the maximum possible $N_{\mathrm{points}}$, in case an inspiral orbit demands indefinitely small $dt$ to capture the final stage before merger. Once two particles merge, the timestep assignment process repeats and the code updates $dt_{\mathrm{new}}$ if needed, ensuring optimal integration performance. 

\subsection{Merger Conditions and Classification: Inspiral Vs Plunge}
\label{sec:merger_cond_class}

In our simulations the condition for a merger event between two BH particles is based on their instantaneous separation \citep[e.g.][]{Merritt_2011,Mannerkoski2021,Rantala2023}. This is defined as the distance of the innermost stable circular orbit (ISCO) around a Schwarzschild black hole with radius 

\begin{equation}
    r_{\mathrm{isco}}=\frac{6GM_{\bullet}}{c^2} = 3 \: R_{\mathrm{s}},
\end{equation}

where $R_{\mathrm{s}}$ is the Schwarzschild radius for an SMBH with mass $M_{\bullet}$. To ensure that PN approximation does not break down at such distances, previous studies set the merger condition to $r \leq 4R_{\mathrm{s}}$ \citep[e.g.][]{Merritt_2011,Mazzolari_2022}. We highlight here that when the distance between an MBH and an SMBH is less than $\sim 6 \: R_{\mathrm{s}} $ the geometry of the spacetime has a strong influence on the orbit dynamics and a proper treatment requires the integration of geodesics coupled with dissipative PN radiation reaction terms \citep{Gair_2006,Sopuerta_2011}. An example of such implementation in $N$-body code has already been tested to study the dynamics and rates of EMRIs in the vicinity of a Schwarzschild or a Kerr SMBH \citep{Amaro_Seoane_2013,Brem_2014}, where radiation reaction terms have been neglected. 
For those reasons we have adopted a merger condition when

\begin{equation}
    r \leq r_{\mathrm{merge}}= 6.5 \: R_{\mathrm{sch}},
\end{equation}

which we have empirically found to be a lower threshold for well-behaved PN-corrected equations of motion for merging binaries in our simulation sample. 

One of the primary goals of our work is to identify and classify different types of black hole merger events: direct plunges and inspirals. The first type of merger orbits arises from perturbations due to relaxation processes and/or strong chaotic encounters ($t_{\mathrm{GW}} > t_{\mathrm{NR}},t_{\mathrm{RR}},t_{\mathrm{ZKL}}$) leading to highly eccentric and radial capture orbits which directly plunge to an SMBH, i.e, when \citep{Merritt_2011,Brem_2014,Amaro_Seoane_2018,Mancieri_2025b} 

\begin{equation}
    r_{\mathrm{peri}}<r_{\mathrm{capt}}=6.5 R_{\mathrm{s}},
\end{equation}

where $r_{\mathrm{peri}}=a(1-e)$ is the pericentre of the orbit and $R_{\mathrm{s}}$ the sum of the Schwarzschild radii of the two black holes from Eq.\eqref{eq:Rsch_comb}. 
For an orbit that barely intersects the capture sphere, the angular momentum and eccentricity are \citep{Merritt_2011},

\begin{equation}
L_{\mathrm{plunge}}=\sqrt{2 G M_{\bullet} r_{\text {capt }}}, \quad e_{\mathrm{plunge}}=1-\frac{r_{\text {capt }}}{a} . 
\end{equation}

Using this definition, an orbit is classified as DP only if $e \geq e_{\mathrm{plunge}}$. More circularized orbits, on the other hand, decouple from external perturbations ($t_{\mathrm{GW}} < t_{\mathrm{NR}}, t_{\mathrm{RR}}, t_{\mathrm{ZKL}}$) and evolve primarily through energy and angular momentum losses due to gravitational-wave emission. These \emph{inspirals} follow a coalescence pathway that is significantly different from that of DPs. Following the above criteria and in line with \cite{Amaro_Seoane_2018, Mazzolari_2022, Mancieri_2025b}, we label as successful inspiral in post-processing, a merger orbit satisfying the following conditions:

\begin{equation}
\left\{\begin{array}{l}
t_{\mathrm{GW}}\left(a, e\right) \leq\left(1-e\right) t_{\mathrm{pert}} \\
e <  e_{\mathrm{plunge}} \\
r_{\mathrm{peri}} \geq r_{\mathrm{capt}} \\
\end{array}\right.
\end{equation}

where $t_{\mathrm{pert}}$ corresponds to any relevant mechanism timescale defined in Section \ref{sec:timescales} and in our case $t_{\mathrm{pert}}=t_{\mathrm{NR}}$.

Slowly evolving orbits, such as gravitational-wave-driven inspirals, are well resolved in the code output using the adaptive timestep scheme (Section~\ref{sec:adaptive_timestep}), allowing their geometric orbital element, especially the geometric eccentricity $e_{\mathrm{g}}$ from Eq.~\eqref{eq:e_g} to be reliably computed. In contrast, the shape of strongly perturbed and rapidly evolving orbits cannot be well described by any set of instantaneous orbital elements. 

\begin{figure}
    \includegraphics[width=0.9\columnwidth]{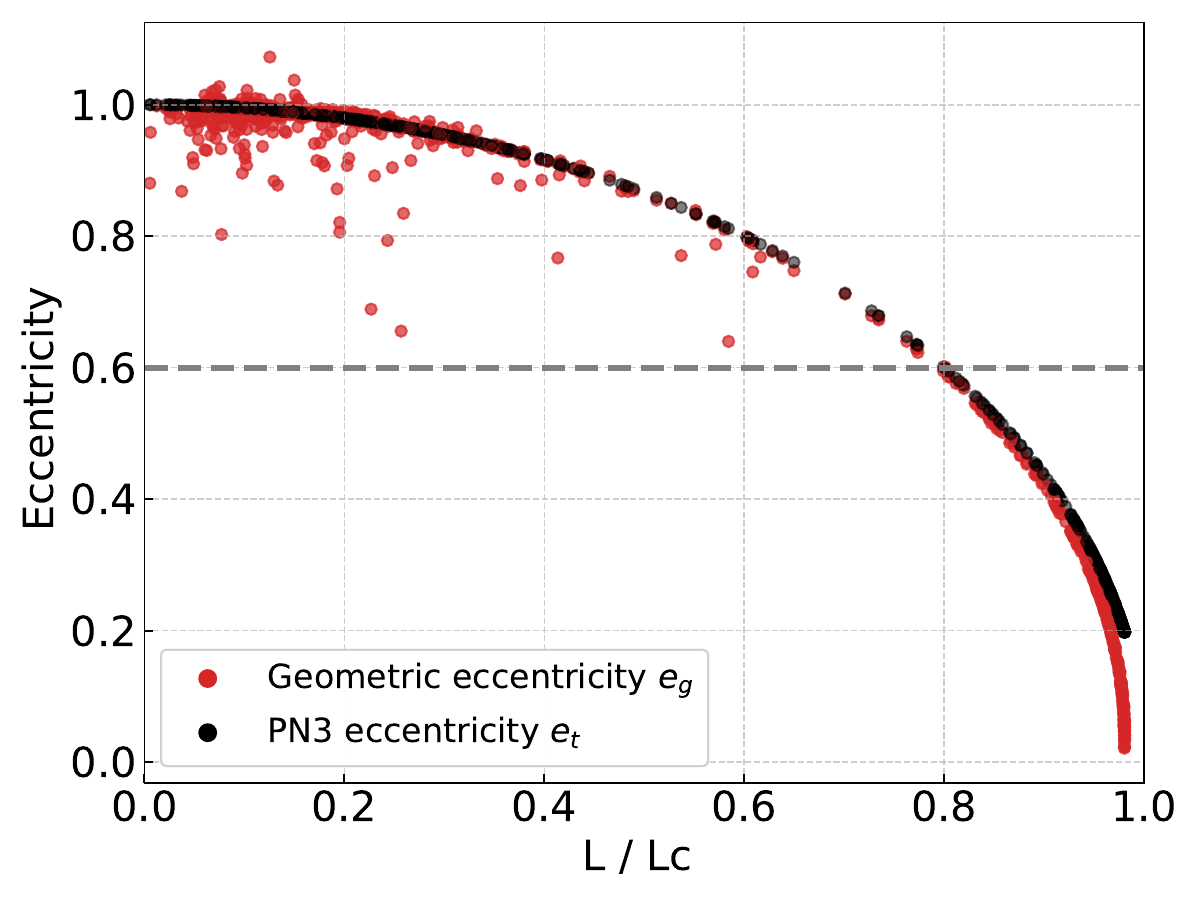}
    \caption{Final orbital eccentricity as a function of normalized angular momentum for all the merger events in our simulations. Different marker colors indicate the two definitions of orbital elements. For orbits with $e < 0.6$, the 3PN-corrected elements overestimate the eccentricity, as the inspirals do not reach $e=0$. Conversely, for highly eccentric orbits ($e \geq 0.6$), the geometric elements fail to accurately represent the orbital properties due to rapid evolution and strong perturbations.}
    \label{fig:L_vs_ecc}
\end{figure}

In Fig.~\ref{fig:L_vs_ecc}, we compare the eccentricities at merger (data extracted from the simulations described in the next Section) obtained from the geometric definition and the 3PN-corrected definition $e_{t}$, plotted as a function of the normalized angular momentum ($L/L_{\mathrm{c}}$), where $L$ corresponds to Eq.~\eqref{eq:EL_PN3} and $L_{\mathrm{c}} = \sqrt{2 G M_{\bullet} a}$ denotes the circular-orbit angular momentum at merger. For slowly evolving inspirals, the geometric definition captures the eccentricity accurately as expected, whereas the 3PN-corrected definition overestimates the final eccentricity and fails to follow the inspiral down to $e=0$. Conversely, for highly eccentric orbits, the geometric elements are unreliable due to resonant relaxation and strong encounters that rapidly modify the orbital shape. For those reasons we adopt a hybrid approach in the analysis of our results, where we use the geometric definition of orbital elements for mergers with $e < 0.6$ and the 3PN-corrected elements for $e \geq 0.6$. Our approach ensures that the shapes of both inspiralling and plunging merger orbits are accurately captured.

\subsection{Initial Conditions and Simulations}
\label{sec:ICS}

\begin{figure*}
	\includegraphics[width=0.49\textwidth, keepaspectratio]{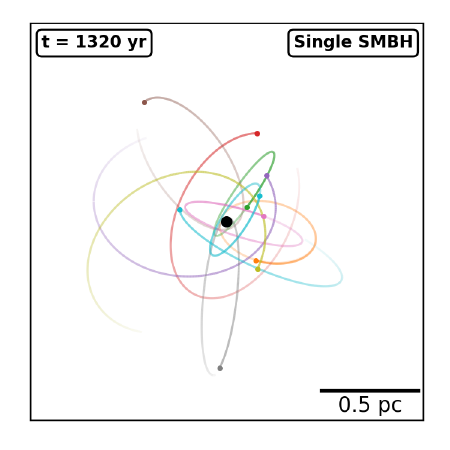}
 \includegraphics[width=0.49 \textwidth, keepaspectratio]{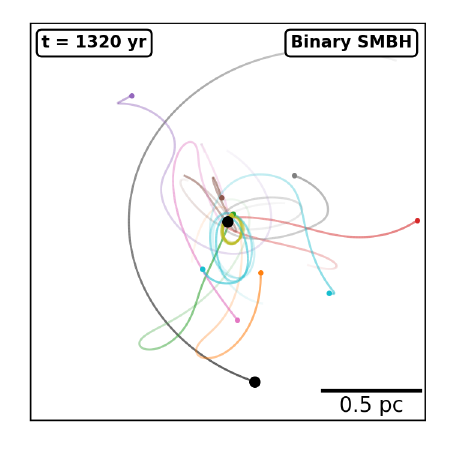}
    \caption{Initial configuration of the systems with wide semi-major axis distribution: $0.05\:\mathrm{pc}  < a \leq 0.5\:\mathrm{pc}$.
    \textbf{Left panel:}  single SMBH setup. \textbf{Right panel:}  system with a companion SMBH on a fixed orbit with $a_{\mathrm{b}}=1 \: \mathrm{pc}$ and $e_{\mathrm{b}}=0.5$ corresponding to an orbital period of $P_{\mathrm{b}}=2000 \: \mathrm{yr}$. Both configurations are shown at $t = 1320 \: \mathrm{yr}$ after the start of the simulations, with trajectory tails tracing the preceding $t = 750 \: \mathrm{yr}$. Notice how the majority of orbits are being disrupted, already after the first pericentre passage of the companion SMBH. The animated versions of these runs are available here: \href{https://doi.org/10.5281/zenodo.17237044}{doi:10.5281/zenodo.17237044}.}
    \label{fig:initconfig}
\end{figure*}

We generate randomly-sampled initial conditions (ICs) for a small cluster of MBHs around an SMBH, where $m= 10^5 \: \mathrm{M_{\odot}}$ and $M_\bullet= 10^9 \: \mathrm{M_{\odot}}$ (left panel in Fig. \ref{fig:initconfig}) . In a second set of simulations, we add a distant equal mass companion SMBH to the primary, on an outer orbit with binary semi-major axis $a_{b}=1 \: \mathrm{pc}$ and eccentricity $e_{\mathrm{b}}=0.5$, corresponding to pericentre passage at $r_{\mathrm{peri}}=0.5 \: \mathrm{pc}$. All the semi-major axes of the MBH orbits are initially within this pericentre distance. We vary the spatial configuration of the MBHs by controlling the lower $a_{\mathrm{min}}$ and upper bound $a_{\mathrm{max}}$ of semi-major axis distribution, such that all MBHs are placed inside a radius of $r=r_{\mathrm{peri}}$, satisfying $a_{\mathrm{max}} \leq r_{\mathrm{peri}}$ (right panel in Fig. \ref{fig:initconfig}) . Our simulations do not account for the presence of a stellar population background; however, we discuss how such an environment would influence the initial dynamical state of our systems in Section \ref{sec:ejections}.

Specifically, we produce ICs for three distributions given in Table~\ref{tab:asemis}.  For each set of simulations (with and without the SMBH companion) , we run 50 realizations for every semi-major axis distribution, resulting in a total of $N_{\mathrm{sim}}=2\times3\times50=300$ simulations. All simulations are run for a total time of $t_{\mathrm{end}}=10 \: \mathrm{Myr}$, so merger events are monitored only up to this timescale.

At coalescence, gravitational waves can impart a recoil \emph{kick} to the remnant BH \citep{Peres_1962,Bekenstein_1973}. In \mstar{} GW recoil kicks are computed following the prescription given in \cite{Zlochower_2015}. The magnitude depends on the binary’s mass ratio and spins, and while our simulations neglect spins—implying only weak recoils—in more general configurations such kicks could eject remnants from their host nuclei with important astrophysical consequences \citep[e.g.,][]{Merritt_2004,Volonteri_2007,Rawlings_2025}.

\begin{table}
\centering
\renewcommand{\arraystretch}{1} 
\begin{adjustbox}{width=0.7\columnwidth}
\begin{tabular}{lc}
\toprule
\textbf{Configuration} & \textbf{Semi-major axis} \\
\midrule
Compact      & $0.005\,\mathrm{pc} < a \leq 0.05\,\mathrm{pc}$ \\
Intermediate & $0.01\,\mathrm{pc}  < a \leq 0.1\,\mathrm{pc}$  \\
Wide         & $0.05\,\mathrm{pc}  < a \leq 0.5\,\mathrm{pc}$  \\
\bottomrule
\end{tabular}
\end{adjustbox}
\caption{Initial MBH configurations based on the distribution of semi-major axis.}
\label{tab:asemis}
\end{table}

\section{Dynamical Evolution of SMBH-MBH systems}
\label{sec:dynamical_evol}

\subsection{Overview}

The initial state of our systems consists of a population of lower-mass MBHs orbiting a central SMBH, with semi-major axes distributed over a wide range ($0.05 \:\mathrm{pc} < a \leq 0.5 \:\mathrm{pc}$). Figure~\ref{fig:initconfig} illustrates this setup for the case of a single SMBH (left) compared to a binary SMBH system (right). In the single-SMBH configuration, the MBHs remain bound in relatively undisturbed orbits around the primary. By contrast, introducing a companion SMBH on a fixed orbit with $a_{\mathrm{b}} = 1 \:\mathrm{pc}$ and $e_{\mathrm{b}} = 0.5$ ($P_{\mathrm{b}} = 2000 \:\mathrm{yr}$) strongly perturbs the system: already by $t = 1320 \:\mathrm{yr}$, following the first pericentre passage ($t\approx 1000 \:\mathrm{yr}$) of the companion, the majority of orbits are disrupted or significantly altered. This demonstrates how the presence of a second SMBH can destabilize the configuration, whereas more compact MBH distributions are more resilient and harder to disrupt.

\begin{figure}
    \includegraphics[width=\columnwidth]{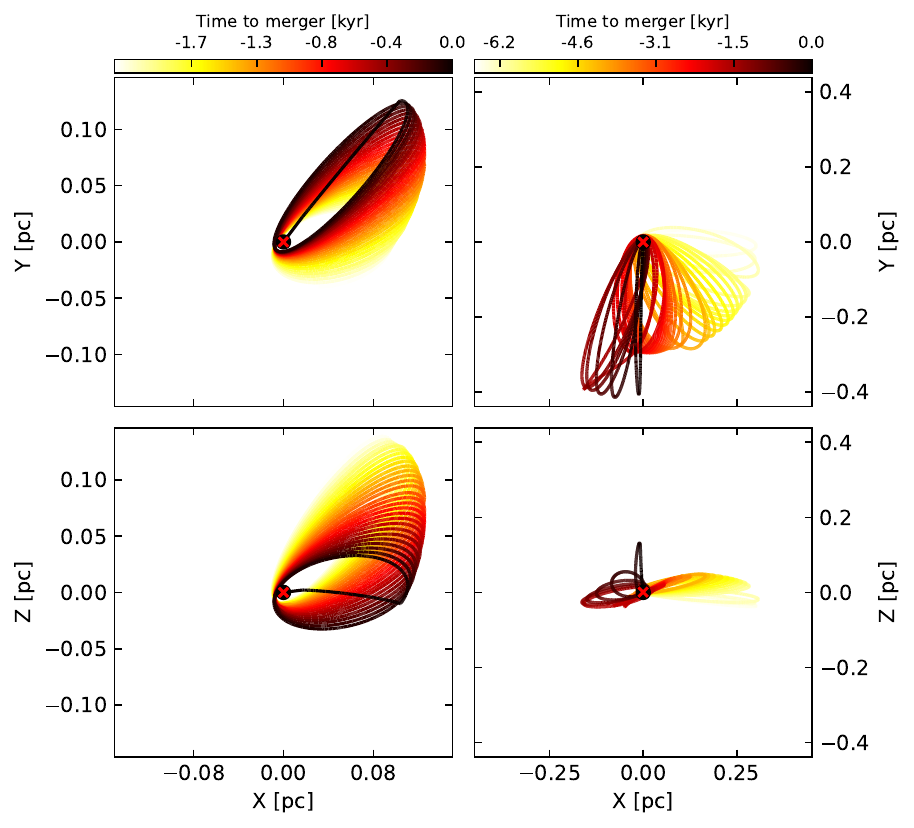}
    \caption{Examples of DP orbits. The colormap traces the time evolution until merger, marked with a red cross. Left panels: single-SMBH run, where a lower-mass MBH precesses due to GR and receives a strong kick from another MBH $\sim 0.5 \: \mathrm{kyr}$ before merger, leading to a rapid plunge. Right panels: binary-SMBH run, where a wider orbit precesses under GR but is perturbed by the companion SMBH $\sim 2 \: \mathrm{kyr}$ before merger, driving it to high eccentricity and eventual plunge. Upper panels show the $x$–$y$ projection, while lower panels display the $x$–$z$ view.}
    \label{fig:dp_orbits}
\end{figure}

\begin{figure}
    \includegraphics[width=\columnwidth]{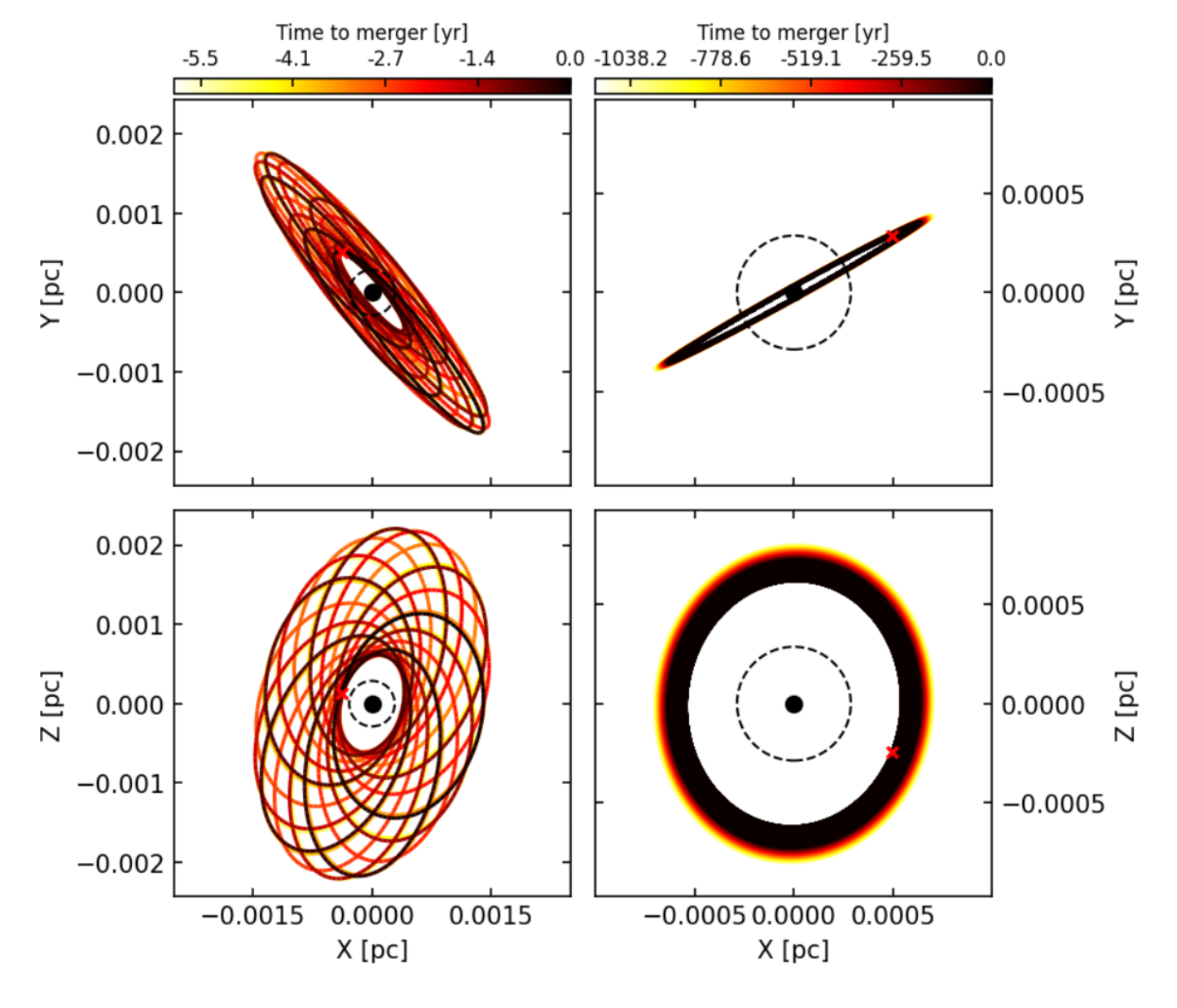}
    \caption{Inspiral example orbits. Left panels: single-SMBH run ($\sim6 \: \mathrm{yr}$ to merger), showing an eccentric orbit that precesses due to relativistic effects while simultaneously shrinking and circularizing via GW emission until $d \leq r_{\mathrm{merge}} = 6.5 \: R_{\mathrm{s}}$. Right panels: binary-SMBH run ($\sim1 : \mathrm{kyr}$ to merger), where the inspiral is driven solely by GW emission. The color map indicates the time to merger (red cross), the dashed circle marks $R_{\mathrm{isco}} = 3 R_{\mathrm{s}}$, and the primary SMBH is shown as a black dot. Upper panels display the $x$–$y$ projection, while lower panels show the $x$–$z$ view.}
    \label{fig:inspiral_orbits}
\end{figure}

In Fig.~\ref{fig:dp_orbits}, we present two representative examples of DP orbits from our simulations. The left panels show a case from a run with a single SMBH, where a lower-mass eccentric MBH orbits the primary while undergoing relativistic precession. The color map indicates the time to plunge. Around $t = 0.5 \: \mathrm{kyr}$ before merger, the orbiting MBH experiences a strong kick from an encounter with another MBH, which drives a rapid plunge onto the primary SMBH. The right panels illustrate an example from a run with a binary SMBH. In this case, the MBH also precesses due to GR, but with a semi-major axis larger by an order of magnitude. At this greater distance, the orbit becomes more susceptible to perturbations from the companion SMBH, which alters its trajectory around $t = 2 \: \mathrm{kyr}$ before merger, exciting a highly eccentric orbit and ultimately leading to the plunge onto the primary.

Figure~\ref{fig:inspiral_orbits} presents examples of inspiral orbits. The left panels, taken from a run with a single SMBH, show a case about $\sim6 \: \mathrm{yr}$ before merger, where relativistic precession and gravitational-wave emission act simultaneously: the eccentric orbit precesses while gradually shrinking and circularizing until it reaches $d \leq r_{\mathrm{merge}} = 6.5 \: R_{\mathrm{sch}}$, our adopted merger criterion. The right panels, corresponding to a run with a binary SMBH, illustrate an inspiral orbit that evolves slowly ($\sim1 \: \mathrm{kyr}$ until merger), driven by GW emission alone, until it reaches $d = r_{\mathrm{merge}}$.

Following the merger criteria and classification outlined in Section~\ref{sec:merger_cond_class}, we present here an overview of the outcomes of our simulations. The total number of mergers, as well as the relative fractions of plunges and inspirals, are summarized in Table~\ref{tab:mergers}. Overall, the presence of a secondary SMBH substantially enhances the merger rate, particularly for wider initial semi-major axis distributions. In these cases, the number of mergers increases by a factor of $\sim 5$ compared to the single-SMBH runs. This enhancement remains significant even for more compact initial configurations, where the merger rate is still elevated by a factor $\sim 2$. All of our merger events involve MBH mergers, with only a small handful of exceptions that have been excluded from the analysis.

\begin{figure*}
    \includegraphics[width=0.8
    \textwidth, keepaspectratio]{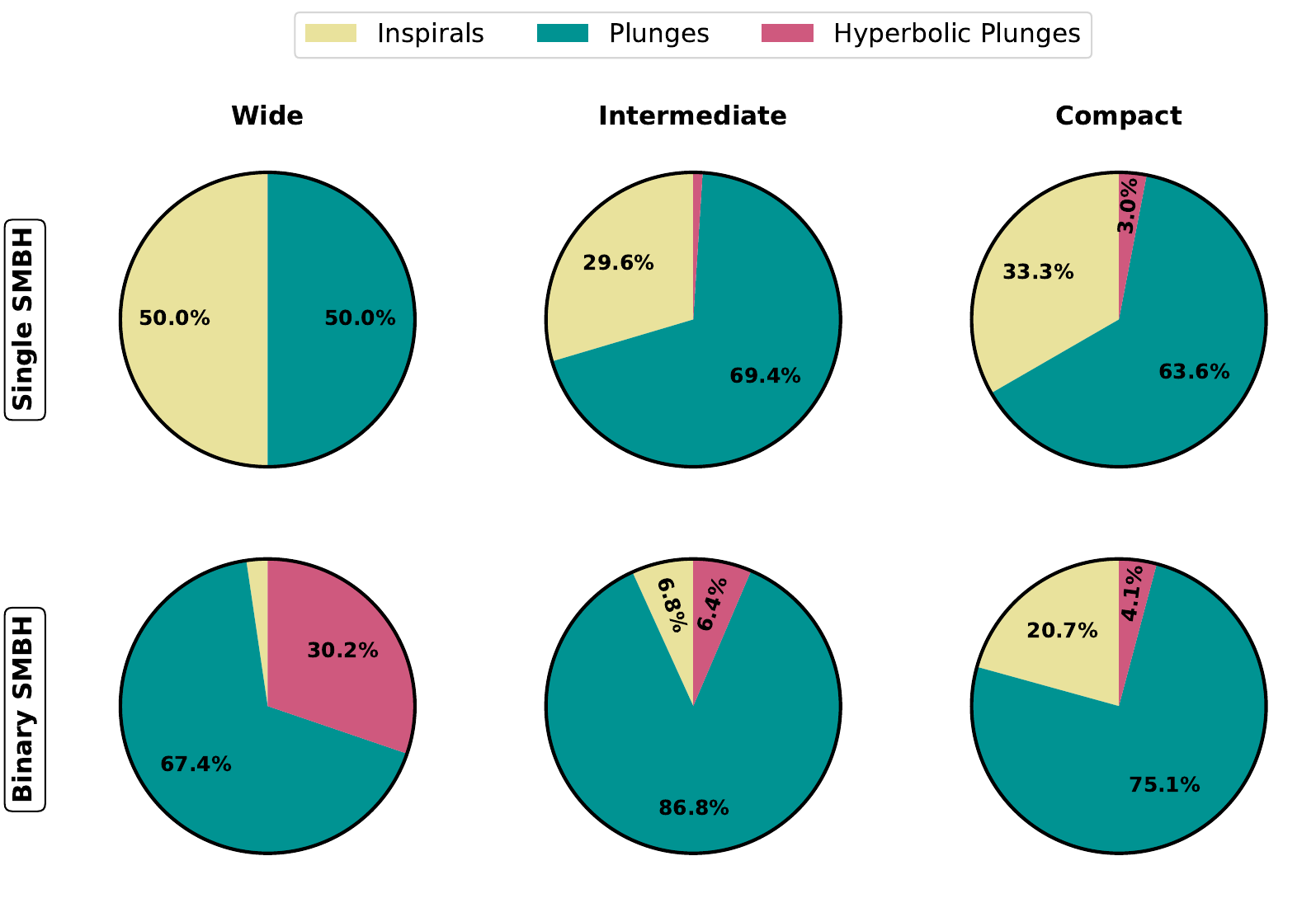}
    \caption{Pie charts illustrating the merger demographics in our simulations. Upper panels show results from single-SMBH runs, while lower panels correspond to binary-SMBH runs. Columns indicate the initial MBH configurations listed in Table~\ref{tab:asemis}. Single-SMBH systems exhibit a substantial fraction of inspirals, whereas the majority of mergers in binary-SMBH runs occur as direct plunges. For the widest MBH configurations, the presence of a companion SMBH produces a significant population of hyperbolic mergers ($e \geq 1$), accounting for $\sim 30\%$ of the total mergers, resulting from strong three-body interactions. In more compact configurations, the fraction of inspirals in binary-SMBH runs increases, reflecting the reduced ability of the secondary SMBH to efficiently disrupt tightly bound MBH systems.}
    \label{fig:merger_pie}
\end{figure*}

For wider initial configurations, the companion SMBH can disrupt the MBH system already during the first pericentre passage at $t \sim 1 , \mathrm{Myr}$, capturing MBHs along its orbit while driving the remaining objects onto highly radial trajectories. This process results either in direct plunges or in the complete ejection of the system (see Section~\ref{sec:ejections}). The impact of this mechanism is evident in Table~\ref{tab:mergers}, where the number of plunges closely tracks the total number of mergers when comparing single- and binary-SMBH runs. We report merger fractions including asymmetric 1$\sigma$ Gehrels confidence intervals \citep{Gehrels_1986}, which are appropriate for Poisson statistics in the low-$N$ regime ($N$ being the number of events) . As expected, the uncertainties become large for bins with very small event counts; a detailed description of the error estimation procedure is provided in Appendix \ref{apndx:A}. In particular, for the two widest configurations, the total plunge fractions in the binary-SMBH case reach $93.2\%$ and $97.7\%$, demonstrating that the overwhelming majority of mergers occur through direct plunges in the presence of a companion SMBH.  Notably as shown in Fig.~\ref{fig:merger_pie}, in the widest configuration nearly $\sim30\%$ of these events are hyperbolic plunges ($e \geq 1$), arising from strong few-body scatterings with the tertiary companion or other MBHs that eject the secondary from the system and drive it directly onto the primary SMBH.

An important outcome of our simulations concerns the relative fraction of inspiral mergers in single- versus binary-SMBH runs. For the wide configurations in the single-SMBH case, although the total number of mergers is lower than in the corresponding binary runs, the fraction of inspirals is substantially higher: $50\%$ and $29.6\%$ for $a_{\mathrm{min}} = 0.05 , \mathrm{pc}$ and $a_{\mathrm{min}} = 0.01 , \mathrm{pc}$, respectively. In contrast, the binary-SMBH runs yield only $2.3\%$ and $6.8\%$ inspirals for the same configurations (lower panels in Fig.~\ref{fig:merger_pie}), highlighting the efficiency of resonant and non-resonant relaxation in gradually driving MBHs towards the GW-dominated regime in the absence of a second SMBH. This result contrasts with previous studies \citep{Naoz_2022, Mazzolari_2022} on less massive SMBH systems, which found that the presence of a secondary SMBH enhances the rate of EMRIs relative to single-SMBH cases. This could be both due to the fact that our systems consist of $10^9 \mathrm{M_{\odot}}$ SMBHs and  $10^5 \mathrm{M_{\odot}}$  MBHs (in contrast to $10^6 \mathrm{M_{\odot}}$ SMBHs surrounded by stellar-mass BHs in the studies above) , as well as the background stellar and BH density and their spatial configuration. In loosely bound MBH systems, which occupy a larger spatial extent, the presence of a secondary SMBH can result in the complete ejection of the MBHs from the system (Section~\ref{sec:ejections}) and potentially from the host galaxy. This ejection mechanism becomes less efficient, producing lower ejection velocities ($V_{\mathrm{ej}}$), for more tightly bound systems (Fig. \ref{fig:ejections}) .
Finally, in the most compact configuration, the relative fractions of inspirals become more balanced: the binary-SMBH runs yield $20.7\%$ inspirals, compared to $33.3\%$ in the single-SMBH case (see Fig.~\ref{fig:merger_pie}), owing to the overall higher number of mergers in the binary runs.

\subsection{Initial State and Merger Pathways}


\begin{table*}
\centering
\renewcommand{\arraystretch}{1.4}
\setlength{\tabcolsep}{5pt}
\begin{adjustbox}{width=\textwidth}
\begin{tabular}{c|cccc|cccc}
\hline
\hline
\multirow{2}{*}{$a$ [pc]} &
\multicolumn{4}{c|}{Single SMBH} &
\multicolumn{4}{c}{Binary SMBH} \\
\cline{2-9}
& Total & IMRIs & Plunges & Hyperbolic Plunges &
  Total & IMRIs & Plunges & Hyperbolic Plunges \\
\hline
0.05 -- 0.5
& 8
& 4 (50.0$^{+35.2}_{-25.4}$\%)
& 4 (50.0$^{+35.2}_{-25.4}$\%)
& 0 (0.0$^{+23.0}_{-0.0}$\%)
& 43
& 1 (2.3$^{+5.4}_{-1.9}$\%)
& 29 (67.4$^{+9.4}_{-8.3}$\%)
& 13 (30.2$^{+8.7}_{-7.4}$\%) \\

0.01 -- 0.1
& 98
& 29 (29.6$^{+6.2}_{-5.5}$\%)
& 68 (69.4$^{+4.9}_{-5.3}$\%)
& 1 (1.0$^{+2.3}_{-0.8}$\%)
& 280
& 19 (6.8$^{+1.9}_{-1.6}$\%)
& 243 (86.8$^{+2.1}_{-2.4}$\%)
& 18 (6.4$^{+1.8}_{-1.5}$\%) \\

0.005 -- 0.05
& 165
& 55 (33.3$^{+5.0}_{-4.4}$\%)
& 105 (63.6$^{+4.0}_{-4.2}$\%)
& 5 (3.0$^{+2.0}_{-1.3}$\%)
& 338
& 70 (20.7$^{+3.0}_{-2.7}$\%)
& 254 (75.1$^{+2.7}_{-3.0}$\%)
& 14 (4.1$^{+1.6}_{-1.3}$\%) \\
\hline
\end{tabular}
\end{adjustbox}
\caption{
Merger outcomes for single and binary SMBH systems under different $a$ distributions.
Each entry lists the number of events and the corresponding fraction of the total,
with asymmetric 1$\sigma$ \protect\cite{Gehrels_1986} confidence intervals.
}
\label{tab:mergers}
\end{table*}

\begin{figure*}
    \includegraphics[width=0.8
    \textwidth, keepaspectratio]{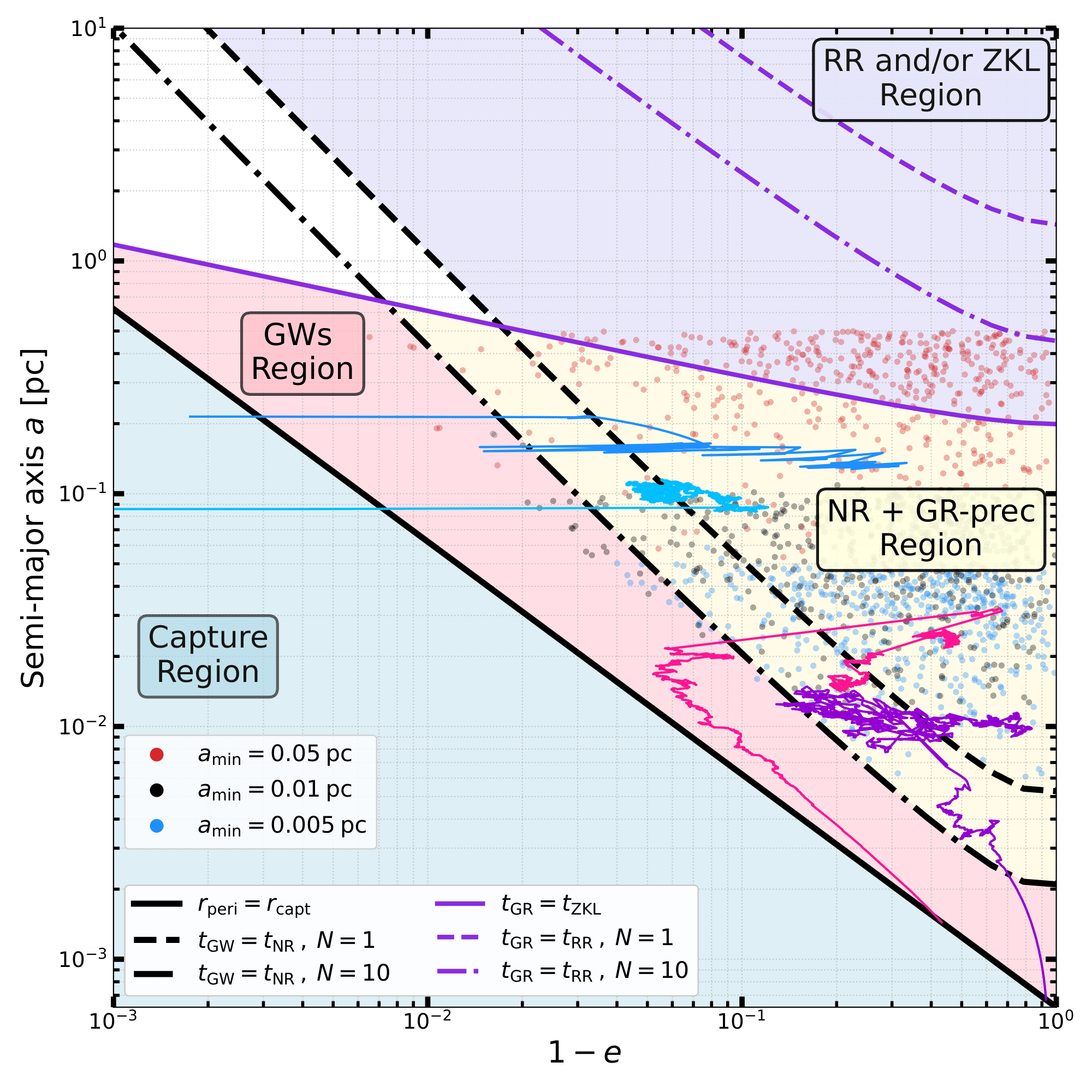}
    \caption{Initial distribution of our simulated systems. The phase space is divided in four different regions due to various key dynamical processes: resonant relaxation (RR) eccentric von Zeipel-Kozai-Lidov (ZKL), non-resonant relaxation (NR), relativistic precession (GR) and GW-driven inspiral (GW). The curves denote the semi-major axes at which the characteristic timescales of two competing processes governing the orbital evolution of a particle are equal. The dot-dashed and dashed purple curves are found by equating $t_{\mathrm{RR}} = t_{\mathrm{GR}}$ and they mark the region (purple) where resonant relaxation is the most rapid process against relativistic precession.  Similarly the black curves correspond to $t_{\mathrm{NR}} = t_{\mathrm{GW}}$ and mark two regions: the pink one where gravitational wave emission dominates the orbital evolution and the yellow where the combined effects of non-resonant relaxation and relativistic precession act on shorter timescales. Different line-styles refer to different number of MBH particles $N$ inside a reference semi-major axis. The black solid curve denotes orbits with $r_{\mathrm{peri}}=r_{\mathrm{capt}}=6.5 \: R{\mathrm{s}}$ marking the region (blue) where a particle in a highly eccentric orbit is captured by and directly plunges to the SMBH. When a companion SMBH with $a_\mathrm{b} = 1$ pc is included, it also induces perturbations on the particles via the ZKL mechanism. To find the critical semi-major axis (solid purple line) above which ZKL dominates over GR, we equate $t_{\mathrm{ZKL}} = t_{\mathrm{GR}}$. Notice how the wide (red) distribution is initially affected by this process. Finally the the light-blue and blue tracks correspond to plunges from a simulation with single and binary SMBH, whereas the pink and purple to inspiral orbits respectively.}
    \label{fig:init_ae}
\end{figure*}

To better understand the early dynamical state of our systems, we begin by discussing the key mechanisms driving the orbital evolution of the MBHs. In Fig.~\ref{fig:init_ae} we present the initial state of our simulations on the $(1-e,a)$ phase space, where different colored markers refer to the three different semi-major axes distributions. The figure is divided in four different regions where different dynamical processes are dominating the evolution of the system. The purple shaded region on the top-right corresponds to the region of phase space where resonant relaxation acts on a shorter timescale than relativistic precession and drives the orbital evolution. The characteristic semi-major axis at which the two processes operate on comparable timescales is obtained by equating Eqs.~\eqref{eq:tRR} and \eqref{eq:tGR} and solving for $a$ ,

\[
\begin{aligned}[t]
t_{\mathrm{GR}} = t_{\mathrm{RR}}
&\;\Rightarrow\; 
\frac{1}{6}(1-e^2)\frac{c^2 a}{G M_\bullet} P = \frac{P}{2\pi} \frac{M_\bullet}{m} \frac{1}{\sqrt{N}} \\
&\;\Rightarrow\; 
a_{\mathrm{crit}}^{\mathrm{RR}} = \frac{3 G M_\bullet^2}{m c^2 \sqrt{N} (1-e^2)},
\end{aligned}
\]

where we have also used Kepler's third law for the orbital period with $m\ll M_\bullet$,
\begin{equation}
\label{eq:kepler}
     P=2 \pi \sqrt{\frac{a^3}{G M_\bullet}}.
\end{equation}

Below $a_{\mathrm{crit}}^{\mathrm{RR}}$ relativistic precession becomes faster and suppresses resonant relaxation. The curve separating those two regimes has been identified and examined in earlier studies \citep{Merritt_2011, Brem_2014, Hamers_2014} and is also known as the \emph{Schwarzschild barrier}. Notice how the curve moves upwards (dot-dashed line for $N=10$ , dashed line for $N=1$) for decreasing MBH number $N$ inside a reference semi-major axis. Since all simulated systems lie below this curve, we can safely neglect resonant relaxation effects in the subsequent analysis. 

The yellow shaded region in Fig.~\ref{fig:init_ae} corresponds to the phase space region where orbital and dynamical evolution are governed by stochastic encounters due to two-body (non-resonant) relaxation coupled with relativistic precession. To separate this region from that where the orbital evolution is dominated by gravitational wave emission, we define another characteristic semi-major axis where the timescale at which non-resonant relaxation operates becomes comparable with GW-driven inspiral timescale.  We obtain this semi-major axis by equating Eqs.~\eqref{eq:tNR} and \eqref{eq:tGW} and again solving for $a$. After some algebra and use of Eqs.~\eqref{eq:F(e)}, \eqref{eq:R(e)} and \eqref{eq:kepler}, we then get 

\[
\begin{aligned}[t]
t_{\mathrm{GW}} = t_{\mathrm{NR}}
&\;\Rightarrow\; 
\frac{5}{64} \frac{c^5 a^4}{G^3 M_\bullet m (M_\bullet + m)} F(e) R(e) = \frac{M_\bullet^2}{m^2 N} P \\
&\;\Rightarrow\; 
a_{\mathrm{crit}}^{\mathrm{NR}} =\left(\frac{128 \pi G^{2.5} M_{\bullet}^{2.5}\left(M_{\bullet}+m\right)}{5 c^5 m N F(e)R(e)}\right)^{2 / 5}.
\end{aligned}
\]

An MBH with $a<a_{\mathrm{crit}}^{\mathrm{NR}}$ is then decoupled from other dynamical perturbations and will evolve towards merger with the SMBH due to energy and angular momentum loses via the emission of gravitational waves alone (pink-shaded region in Fig.~\ref{fig:init_ae}). We show the effect of particle number inside some reference semi-major axis with different line-styles, where again for decreasing $N$ the curve moves upwards. This curve serves as a key diagnostic, delineating whether an orbit’s evolution is dominated by stochastic perturbations—such as non-resonant relaxation and chaotic encounters—or by the gradual, decoupled inspiral driven by gravitational wave emission. It has been employed in previous studies to distinguish between direct plunges into the SMBH and slow inspirals that can lead to the formation of EMRIs \citep{Merritt_2011,Brem_2014,Hamers_2014,Mancieri_2025b}. 

In addition, strong scatterings \citep[see e.g.,][]{Teboul_2025} with a tertiary SMBH or chaotic three-body interactions can drive MBHs onto hyperbolic trajectories ($e \geq 1$), resulting in high-velocity encounters or ejections from the system. Finally, if any of the above processes drive a particle to an extremely (almost radial) eccentric orbit, then it can by-pass the GW-dominated region and be captured by the SMBH when $r_{\mathrm{peri}}<r_{\mathrm{capt}}=6.5 \: R{\mathrm{s}}$. The corresponding semi-major axis limit below which such direct capture occurs (blue-shaded region in Fig.~\ref{fig:init_ae}) is then 

\begin{equation}
    r_{\mathrm{peri}}=r_{\mathrm{capt}} \Rightarrow a(1-e)=r_{\mathrm{capt}}\Rightarrow a_{\mathrm{capt}} = \frac{r_{\mathrm{capt}}}{1-e}.
\end{equation}

It is important to mention at this stage, that the inclusion of a companion SMBH can alter significantly the early dynamical evolution of the system-especially for wide MBH configurations- causing eccentricity oscillations to an inner binary consisting of the primary SMBH and an MBH via the eccentric ZKL mechanism \citep{Bode_2013,Naoz_2022,Mazzolari_2022}. The solid purple curve in Fig.~\ref{fig:init_ae} corresponds to the semi-major axis where the timescales between ZKL and GR-precession become comparable and is obtained by equating this time Eqs.~\eqref{eq:tGR} and \eqref{eq:tZKL}. In the region above this line, ZKL dominates the early evolution of an MBH in the presence of the secondary SMBH.

Additionally, in Fig. \ref{fig:init_ae} the evolutionary tracks for two plunge and two inspiral orbits are plotted. The light-blue line is an example of a plunging orbit in a simulation with a single SMBH, which initially evolves due to the cumulative effects of NR until it receives a strong perturbation enough to drive the MBH deep inside the capture radius. Similarly, the blue line corresponds to a plunging orbit in the presence of the second SMBH, where the MBH undergoes a 3-body encounter, boosting its eccentricity and resulting in a radial orbit crossing the capture radius. These tracks are representative of direct plunge events, and the majority of such events in our simulations follow similar trajectories, exhibiting a rapid transition from mildly eccentric orbits to highly radial plunges into the SMBH.

In contrast, the pink track illustrates an inspiral orbit in a simulation with a single SMBH, where the early evolution is driven by non-resonant relaxation and GR precession. The MBH eventually enters the GW-dominated region and gradually inspirals toward merger via GW emission. The purple track shows a similar case in a binary SMBH simulation, but here the MBH enters the GW-dominated regime with low eccentricity and semi-major axis, producing an almost circular inspiral ($e \sim 0$). Inspiral tracks are generally more sensitive to the initial location in phase space (Fig.~\ref{fig:init_ae}): depending on which region they enter from—whether dominated by NR, RR, or ZKL mechanisms—the resulting inspiral can exhibit substantially different semi-major axes and eccentricities at the onset of GW-driven evolution. Most inspiral events follow trajectories qualitatively similar to these examples, but their detailed properties are strongly influenced by the dominant dynamical processes in the region from which they enter the GW-dominated regime.

\begin{flushleft}
\begin{alignat*}{2}
t_{\mathrm{GR}} &= t_{\mathrm{ZKL}} \\
&\Rightarrow \frac{1}{6}(1-e_{\mathrm{in}}^2)\frac{c^2 a}{G M_1} P_{\mathrm{in}} = \frac{16}{30 \pi} \frac{M_{1}+M_{2}+m}{M_{2}} \frac{P_{\mathrm{out}}^2}{P_{\mathrm{in}}}\left(1-e_{\mathrm{out}}^2\right)^{3 / 2} \\
&\Rightarrow a_{\mathrm{crit}}^{\mathrm{ZKL}} =\left[\frac{16}{5 \pi} \frac{G M_1\left(M_1+m\right)}{c^2} \frac{\left(1-e_{\mathrm{out}}^2\right)^{3 / 2}}{M_2}  \frac{a_{\mathrm{out}}^3}{1-e_{\mathrm{in}}^2}\right]^{1 / 4} ,
\end{alignat*}
\end{flushleft}

where $M_{1}$ and $M_{2}$ are the masses of primary and secondary SMBH and $P_{\mathrm{in}}$, $e_{\mathrm{in}}$ are the period and eccentricity of the inner binary. The outer binary is on a fixed orbit (Section \ref{sec:ICS}) with $a_{\mathrm{out}}=1\mathrm{pc}$ and $e_{\mathrm{out}}=0.5$. This separating curve is essentially the \emph{Schwarzschild barrier} for SMBH binaries \citep{Bode_2013}. The key difference for the single SMBH simulations is that RR has no influence on the early dynamical evolution of the system and is already suppressed by GR-precession, whereas in the binary SMBH runs, the ZKL mechanism has a significant influence on the wide (red markers) configuration.

In general, eccentricity growth induced by the ZKL mechanism can be significantly suppressed by general relativistic precession \citep{Bonetti_2016,Bonetti_2019, Lim_2020}. Moreover, given that the evolution of our $N=10$ systems is inherently chaotic, the hierarchical structure of the triple configuration will eventually be disrupted, rendering the secular ZKL formalism inapplicable. In the context of chaotic three-body dynamics, one may define a critical semi-major axis, $a_{\mathrm{chaos}}$, for the outer binary, which delineates the boundary between distinct dynamical regimes \citep{Mardling_2001, Vynatheya_2022}. A detailed investigation of these regimes requires alternative methodologies, which lie beyond the scope of the present work.

\begin{figure}
    \includegraphics[width=\columnwidth]{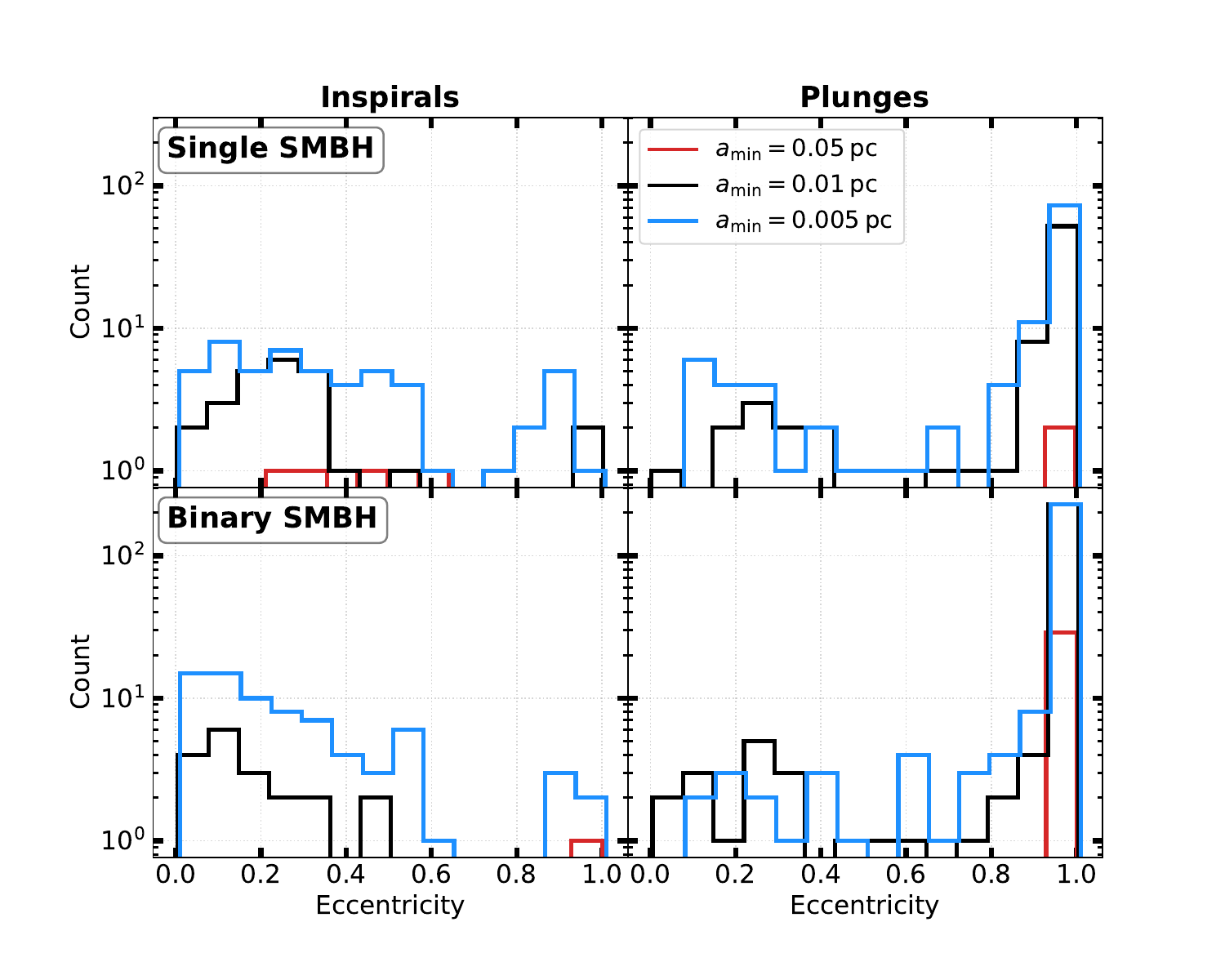}
    \caption{Distribution of final (at merger) eccentricity of direct plunges (right) and inspirals (left). The eccentricity for DPs peaks at $e\geq0.9$, as expected for the majority of such events. MBHs on near-circular orbits with $a\gtrsim r_{\mathrm{capt}}$, can directly plunge to the SMBH, after receiving a perturbation from an another MBH, which explains the low-eccentricity DPs on the right panels. For IMRIs, there is not preferred eccentricity from single SMBH runs, with the majority of such events having $e\leq0.6$ when a second SMBH is included. MBH orbits with small semi-major axes ($a \approx 0.01 \mathrm{pc}$) and large eccentricities can still reach the GW-dominated region of phase space, where they evolve into short-lived inspirals that merge while retaining high eccentricity. }
    \label{fig:ecc_distr}
\end{figure}

\subsubsection{The influence of the stellar background on the dynamical evolution}

In our simulations and analysis we did not include a stellar background distribution. Here, we briefly assess how the relevant relaxation timescales would be affected in a typical massive early-type galaxy (ETG). We a typical ETG model (e.g. \citealt{Rantala2018,Rantala_2024}) characterized by a central density of $\rho_{\mathrm{c}}=10^4 \: \mathrm{M_{\odot} pc^{-3}}$ at $r=10$ pc and a Hernquist density profile \citep{Hernquist_1990} with an inner power-law slope $\rho(r)\propto r^{-\gamma}$, where $\gamma=1$. For an average stellar mass of $m\approx0.4 \:\mathrm{M_{\odot}}$ from an evolved stellar population within a radius of $r=1\mathrm{pc}$, this yields a total cumulative stellar mass $M_\star(r<1\mathrm{pc})=4.19\times10^4 \: \mathrm{M_{\odot}}$ and a total number of stars $N_\star(r<1\mathrm{pc})=1.05\times10^5$. Thus, the SMBH dominates the mass budget of the central parsec of the system the entire MBH system is well embedded within the influence radius of the primary SMBH.

Considering an MBH orbiting a supermassive black hole with a semi-major axis of $a = 0.01 \: \mathrm{pc}$ and $e=0.5$ (same as the example orbit used for the timescales in Table~\ref{tab:timescale}) , the expected number of stars in the above early-type galaxy  within that radius is $N_\star(r < 0.01\mathrm{pc}) \sim 0$. Consequently, both resonant relaxation (RR) and non-resonant relaxation (NR) from the stellar component play only a minor role in the dynamics at such small scales. For the same stellar distribution but with an MBH on a wider orbit ($a = 0.5 \: \mathrm{pc}$), the enclosed number of stars rises to $N_\star(r < 0.5\mathrm{pc}) \sim 5000$. At these distances, the corresponding relaxation timescales are $t_{\mathrm{RR}} \approx 1.45 \times 10^4  \mathrm{yr}$ and $t_{\mathrm{NR}} \approx 8 \times 10^6 \:  \mathrm{yr}$.  Relativistic precession weakens with increasing separation, yielding $t_{\mathrm{GR}} \approx 1.37 \times 10^6 \: \mathrm{yr}$. Since $t_{\mathrm{RR}} < t_{\mathrm{GR}}$, relativistic precession cannot efficiently suppress RR at large separations, which therefore becomes the dominant and fastest mechanism governing orbital evolution. At those distances, for a subsystem of $N = 10$ MBHs, the RR and NR timescales are $t_{\mathrm{RR}} \approx 5.27 \times 10^5 \: \mathrm{yr}$ and $t_{\mathrm{NR}} \approx 1.05 \times 10^{10} \: \mathrm{yr}$, respectively, further confirming that RR governs the evolution of the MBH ensemble. Thus the presence of a stellar background would only enhance the overall efficiency of RR-driven evolution.

This highlights the importance of including a stellar background—particularly at larger orbital scales where GR effects are weaker and the MBH may initially reside—as the stellar torques can substantially enhance angular-momentum diffusion. In this regime, RR dominates over both NR and GR, while NR and GR act on comparable timescales ($t_{\mathrm{NR}} \sim t_{\mathrm{GR}}$) and thus compete. By contrast, the mass-precession timescale for the stellar background ($t_{\mathrm{MP}} \approx 1.15 \times 10^8 \: \mathrm{yr}$) remains orders of magnitude longer than $t_{\mathrm{GR}}$, confirming that relativistic precession continues to limit its dynamical influence even at these separations. For an $N = 10$ MBH subsystem, however, $t_{\mathrm{MP}} \approx 6.05 \times 10^5 \: \mathrm{yr}$, bringing it into direct competition with RR and further emphasizing the complex interplay of relaxation processes in the multi-MBH regime.

\subsection{Merger eccentricity distribution}

We focus on the eccentricity distribution of our merger events, as it encodes critical information about the dynamical environment in which the coalescence occurs. In particular, large residual eccentricities at merger can serve as a diagnostic of triple-MBH interactions, which can drive binaries to extreme values of $e \gtrsim 0.9$ and accelerate coalescence, requiring an accurate relativistic treatment \citep[e.g.,][]{Bonetti_2016}. Recent work further demonstrates that high-eccentricity mergers are a natural outcome of chaotic three-body encounters among MBHs in galactic nuclei \citep{Bonetti_2019}. Such signatures are especially relevant for LISA, since residual eccentricity leaves an imprint on their GW-waveform. Detecting eccentric mergers would thus not only provide evidence for triple-MBH interactions but also offer a unique probe of the dynamical assembly channels of MBHs in galactic centers.

Figure~\ref{fig:ecc_distr} shows the eccentricity distribution from our simulations measured at the time of merger. The right panels correspond to direct plunge (DP) events, with the expected peak at $e \geq 0.9$ in both single- and binary-SMBH cases. MBHs on near-circular orbits with $a \gtrsim r_{\mathrm{capt}}$ can still undergo direct plunges if perturbed by another MBH, which explains the presence of low-eccentricity DPs in the right panels. We do not observe significant differences between single- or binary-SMBH runs, nor a strong correlation with the compactness of the system. The broader (red) distribution from the binary-SMBH runs (bottom panel) is dominated by such highly eccentric mergers triggered by encounters with the secondary SMBH.

The eccentricity distribution of inspirals is presented in the left panels of Fig.~\ref{fig:ecc_distr}. Most inspirals occur with $e < 0.5$, consistent with orbital evolution primarily driven by GW emission. For IMRIs, there is no preferred eccentricity in the single-SMBH runs, whereas in the binary-SMBH case the majority of events have $e \leq 0.6$. A small subset with $e \gtrsim 0.6$ corresponds to MBHs that enter the GW-dominated regime (highlighted in Fig.~\ref{fig:init_ae}) on highly eccentric orbits, often as a result of chaotic three- or four-body interactions. These produce short-lived inspirals that merge while retaining substantial eccentricity. In particular, MBH orbits with small semi-major axes ($a \approx 0.01 ,\mathrm{pc}$) and large eccentricities can evolve directly into this regime, leading to high-$e$ mergers.

Finally, if the number of particles within the orbit of an MBH is sufficiently large (in our setup, $N_{\mathrm{max}} = 10$), two-body relaxation (NR) could, in principle, drive eccentricities to large values on longer timescales; however, this effect is not observed in our simulations. The limited number of inspirals in the wider (red) MBH configuration further prevents us from drawing robust conclusions for those systems.

\subsection{Ejections from Single and Binary SMBHs}
\label{sec:ejections}

\begin{figure}
    \includegraphics[width=\columnwidth]{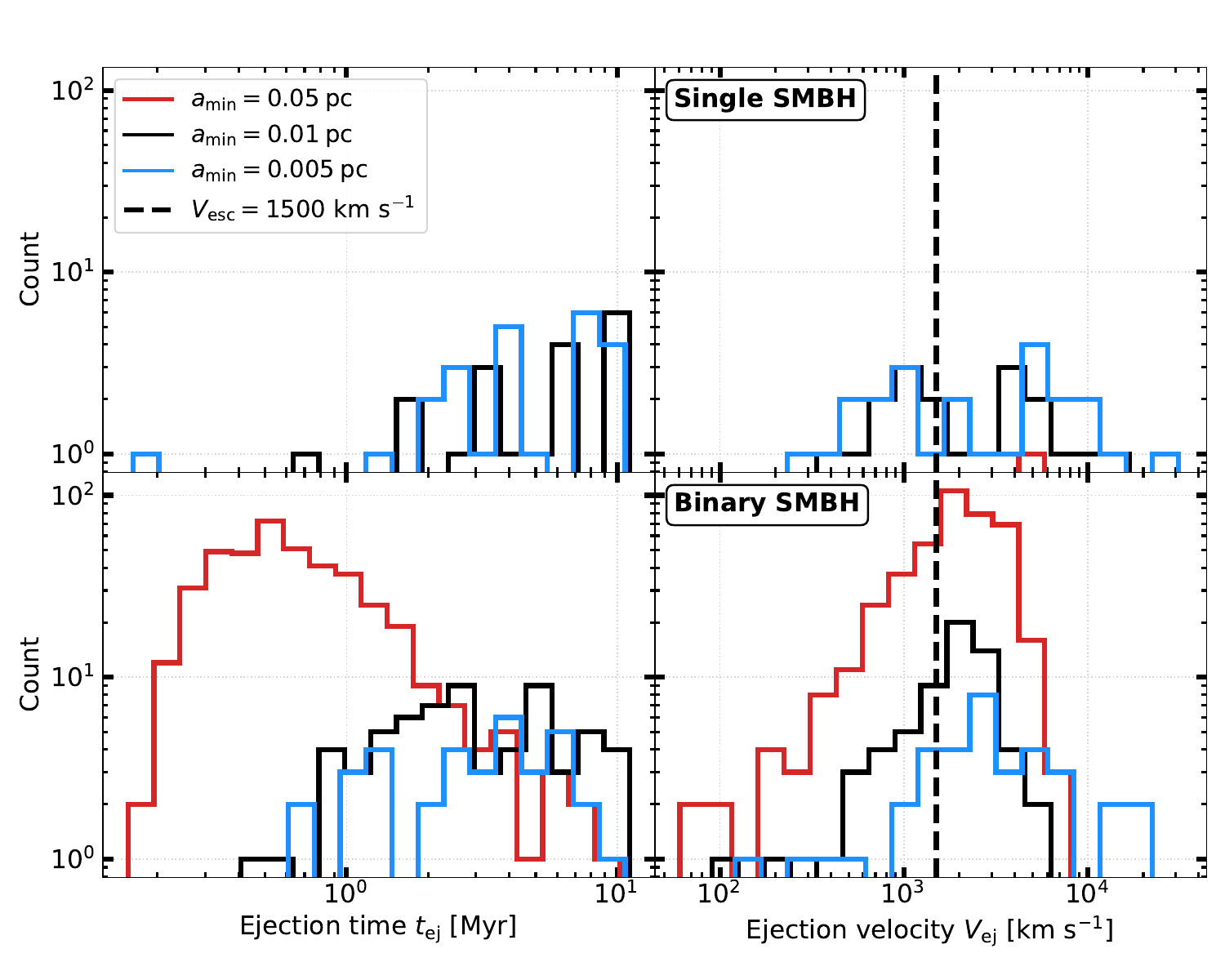}
    \caption{Distribution (top panel: single SMBH runs, bottom
panel: binary SMBH runs) of ejection times $t_{\mathrm{ej}}$ (left) and ejection velocities (right) of MBHs which have escaped in the runs. The vertical black dashed line correspond to the escape velocity of a typical ETG, where $V_{\mathrm{esc}}=1500 \mathrm{~km} \mathrm{~s}^{-1}$.}
    \label{fig:ejections}
\end{figure}

In our simulations, particles that become unbound are removed from the system once they reach a distance of $r_{\mathrm{esc}}=1 \: \mathrm{kpc}$. Figure~\ref{fig:ejections} shows the distributions of ejection times, $t_{\mathrm{ej}}$ (left panels), and ejection velocities, $V_{\mathrm{ej}}$ (right panels), for the escaped MBHs, measured at $r_{\mathrm{esc}}$. 

When MBH–SMBH binaries become sufficiently hard \citep{Quinlan_1996,Sesana_2006}, they can efficiently eject stars and COs that pass nearby. This dynamical slingshot mechanism \citep{Hills_1988} has been widely studied in the context of inspiralling star clusters hosting IMBHs, where an SMBH–IMBH binary can accelerate stars to high velocities \citep{Baumgardt_2006,Sesana_2007,Sesana_2009a,Evans_2023,Rasskazov_2019}. In merging star clusters hosting MBHs, \citet{Souvaitzis_2025} further demonstrated that hard MBH binaries not only increase the number of ejected stars and compact objects but also significantly boost their ejection velocities, highlighting the efficiency of such dynamical interactions. A related mechanism arises when a stalled MBH binary interacts with a third incoming MBH \citep{Ryu_2018}, forming a transient triplet: chaotic three-body encounters can both harden the surviving binary and eject the lighter component at high velocity \citep{Mikkola_1990}. These processes underscore the role of few-body dynamics in shaping both the hardening of MBH binaries and the population of wandering or ejected MBHs in galactic nuclei.

In our results, the presence of a second SMBH dramatically enhances the ejection rate for less bound MBH configurations (wide distribution shown in red). All ejections in our simulations arise from purely gravitational interactions—both Newtonian and relativistic in nature—and are not associated with gravitational-wave recoil kicks. The peak of the red $t_{\mathrm{ej}}$ distribution occurs at $t_{\mathrm{ej}} \sim 0.5 \: \mathrm{Myr}$, which roughly corresponds to the typical fly-by time associated with the most common ejection velocities ($V_{\mathrm{ej}}\approx2000 \mathrm{~km} \mathrm{~s}^{-1}$). As a result, many of the escaping MBHs are expelled from the central system during the initial disruption phase, when the secondary SMBH first perturbs the configuration. This leads to a substantial excess of high-velocity ejections, with 283 out of 419 MBHs removed from the host galaxy due to these interactions.

The effect of the second SMBH in tightly bound systems is weaker, while they still increase the total number of ejections compared to runs with a single SMBH: from 12 (out of 19) to 44 (out of 64) for the intermediate configuration (black histograms) and from 15 (24) to 27 (out of 36) for the compact one (blue histograms). The effect of the compactness of the system is clearly visible in the $V_{\mathrm{ej}}$ distribution (bottom-right panel in Fig. \ref{fig:ejections}), except from a few extreme cases in the compact configuration leading to $V_{\mathrm{ej}}>10^4 \mathrm{~km} \mathrm{~s}^{-1}$.

\section{Gravitational Waves}
\label{sec:GWs}

A binary system on a circular orbit emits gravitational waves at a frequency $f_{\mathrm{GW}}$ which is simply twice the orbital frequency \citep[e.g.][]{Blanchet2006,Maggiore2007} or

\begin{equation}
    f_{\mathrm{GW}} = 2 f_{\mathrm{orb}} = 2 / P,
\end{equation}

where $P$ is the orbital period Eq.\eqref{eq:kepler}. However, for eccentric binaries, the GW emission is strongly concentrated near pericenter, resulting in a shift of the peak frequency towards higher harmonics. In accordance with the foundational work of \cite{Peters1963}, which utilized the lowest-order (1PN) post-Newtonian approximation, it was demonstrated that the power of the $n$-th harmonic, with frequency $f_{\mathrm{GW,n}}=nf_{\mathrm{orb}}$ (with integer $n \geq1$), is described by  

\begin{equation}
    P_\mathrm{n} = \frac{32}{5} \frac{G^4}{c^5} \frac{\mathcal{M}^5}{a^5} g(n, e),
\end{equation}

where $\mathcal{M}$ is the chirp mass,

\begin{equation}
\label{eq:chirpm}
\mathcal{M}=\frac{(M_\bullet m)^{3 / 5}}{(M_\bullet+m)^{1 / 5}}.
\end{equation}

The dimensionless function $g(n,e)$ quantifies the excess power emitted in the $n-$th harmonic relative to a circular orbit and is given by \citep{Peters1963} 

\begin{equation}
\label{eq:g_ne}
\begin{aligned}
& g(n, e)=\frac{n^4}{32}\left\{\left[J_{n-2}(n e)-2 e J_{n-1}(n e)+\frac{2}{n} J_n(n e)\right.\right. \\
& \left.\quad+2 e J_{n+1}(n e)-J_{n+2}(n e)\right]^2+\left(1-e^2\right)\left[J_{n-2}(n e)\right. \\
& \left.\left.\quad-2 J_n(n e)+J_{n+2}(n e)\right]^2+\frac{4}{3 n^2} J_n(n e)^2\right\},
\end{aligned}
\end{equation}

where $e$ is the orbital eccentricity and $J_n$ are Bessel functions of the first kind. 

\subsection{Peak Gravitational Wave Frequency of Merger Events}

\begin{figure}
    \includegraphics[width=\columnwidth]{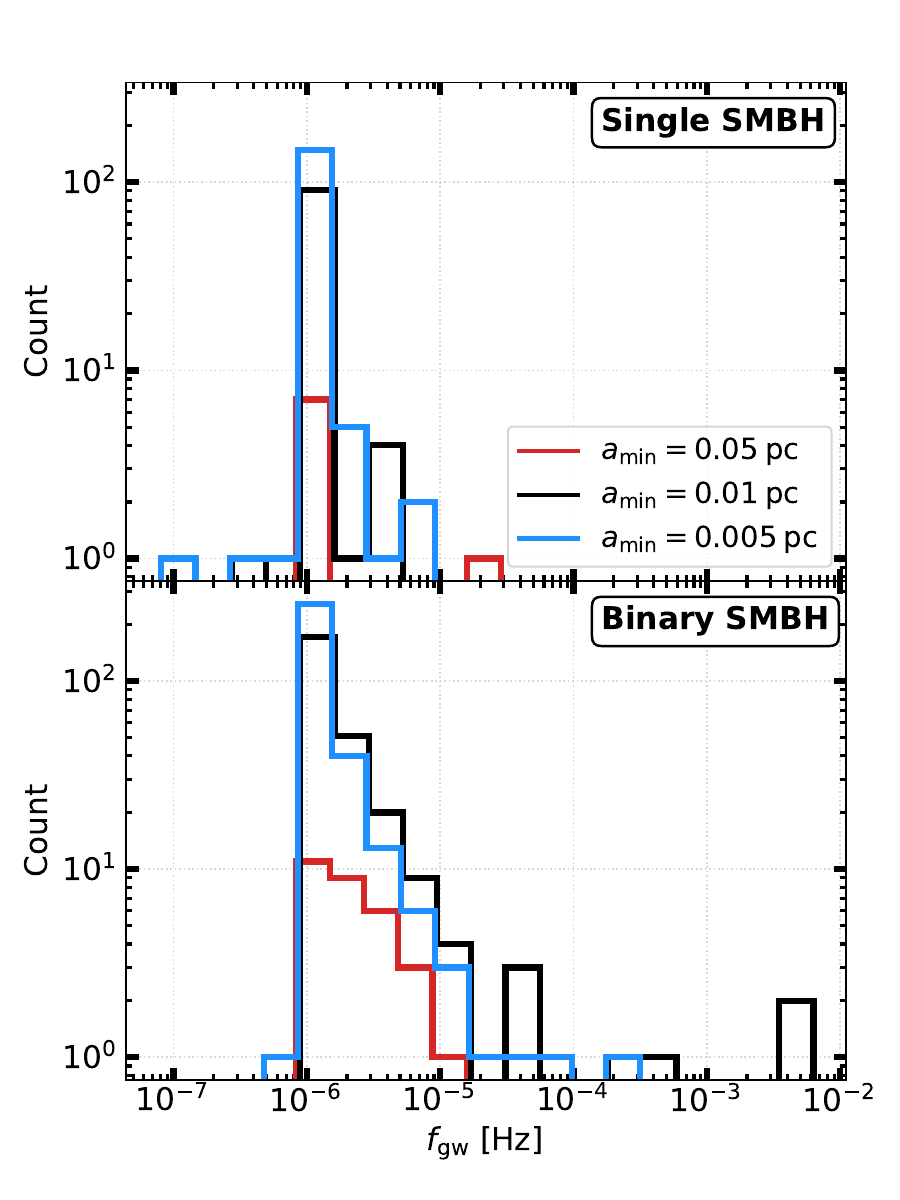}
    \caption{Distribution of the peak gravitational wave frequency $f_{\mathrm{GW}}^{\mathrm{peak}}$ for the merged systems in our simulations (top panel: single SMBH runs, bottom panel: binary SMBH runs). The presence of a companion SMBH systematically broadens the range of gravitational-wave frequencies at which mergers occur. Additionally, the compactness or spatial extent of the initial semi-major axis distribution (indicated by the different colors) further influences the spread of the frequency distribution, while also increasing the total number of merger events.}
    \label{fig:f_peak}
\end{figure}

The characteristic chirp mass of our binaries, with a primary SMBH of $M_\bullet = 10^9  M_\odot$ and a secondary MBH of $m = 10^5  M_\odot$, is $\mathcal{M}\simeq 4\times10^6 \: \mathrm{M_{\odot}}$ computed via Eq.~\eqref{eq:chirpm} . We expect LISA to detect MBH binaries with $\mathcal{M}\sim 10^4-10^7 \: \mathrm{M_{\odot}}$ \citep{Klein_2016,Amaro_Seoane_2017_LISA}, placing our system on that range, while remaining below the typical sensitivity threshold for PTAs $\mathcal{M}\sim 10^8-10^{10} \: \mathrm{M_{\odot}}$  \citep{Hobbs_2010,Sesana_2013}. Consequently, while the intrinsic chirp mass of our binaries falls within LISA’s nominal sensitivity range, the redshift and mass ratio can shift the observed GW frequencies. We discuss the implications for their detectability with LISA and PTAs in the next section (Section \ref{sec:strains}).

To check whether our sources enter the frequency band of PTAs and/or LISA, we compute the distribution of the emitted GW frequencies. Since a large fraction of our sources are eccentric, it is important to make sure we account for the correct harmonics contribution for each case. This corresponds to the value of $n_{\mathrm{peak}}(e)$ for which Eq. \eqref{eq:g_ne} is maximum. The peak GW frequency then becomes 

\begin{equation}
\label{eq:f_peak}
f_{\mathrm{GW}}^{\mathrm{peak}}(e)=n_{\mathrm{peak}}(e) f_{\mathrm{orb}} .
\end{equation}

\cite{Wen_2003} provides a fitting function for $n_{\mathrm{peak}}(e)$, a prescription that has been widely adopted in studies of eccentric compact binary sources, \citep[see e.g.][]{Brown_2010,Gold_2013,Samsing_2014} capturing the frequency shift induced by eccentricity while remaining computationally efficient for large ensembles of binaries. However, \cite{Hamers_2021} have found that this fit behaves poorly for less eccentric sources $(e \lesssim 0.8)$ and provided a corrected version of the fit capturing the full eccentricity range:

\begin{equation}
n_{\mathrm{peak}}(e) \simeq 2\left(1+\sum_{k=1}^4 c_k e^k\right)\left(1-e^2\right)^{-3 / 2},
\end{equation}

where $c_1 = -1.01678$, $c_2 = 5.57372$, $c_3 = -4.9271$, and $c_4 = 1.68506$. Using this approach, we can robustly estimate the GW signal from each merger, including highly eccentric systems, and construct the corresponding GW spectrum relevant for PTA experiments as well as LISA-related sources.

In Fig.~\ref{fig:f_peak} we present the distributions of the peak gravitational wave frequency of mergers from the simulations with a single (top panel) or a binary SMBH (bottom panel) . In all of our simulations we find the that the presence of the companion SMBH systematically broadens the frequency distribution pushing it towards higher values. This is essentially due to the excess of highly eccentric plunging orbits (right panels in Fig.~\ref{fig:ecc_distr}), where the number of higher harmonics $n_{\mathrm{peak}}$ can be extremely large leading to increased peak gravitational wave frequencies $f_{\mathrm{GW}}$ at merger.

\subsection{Characteristic Strain}
\label{sec:strains}

\begin{figure*}
    \includegraphics[width=1\textwidth, keepaspectratio]{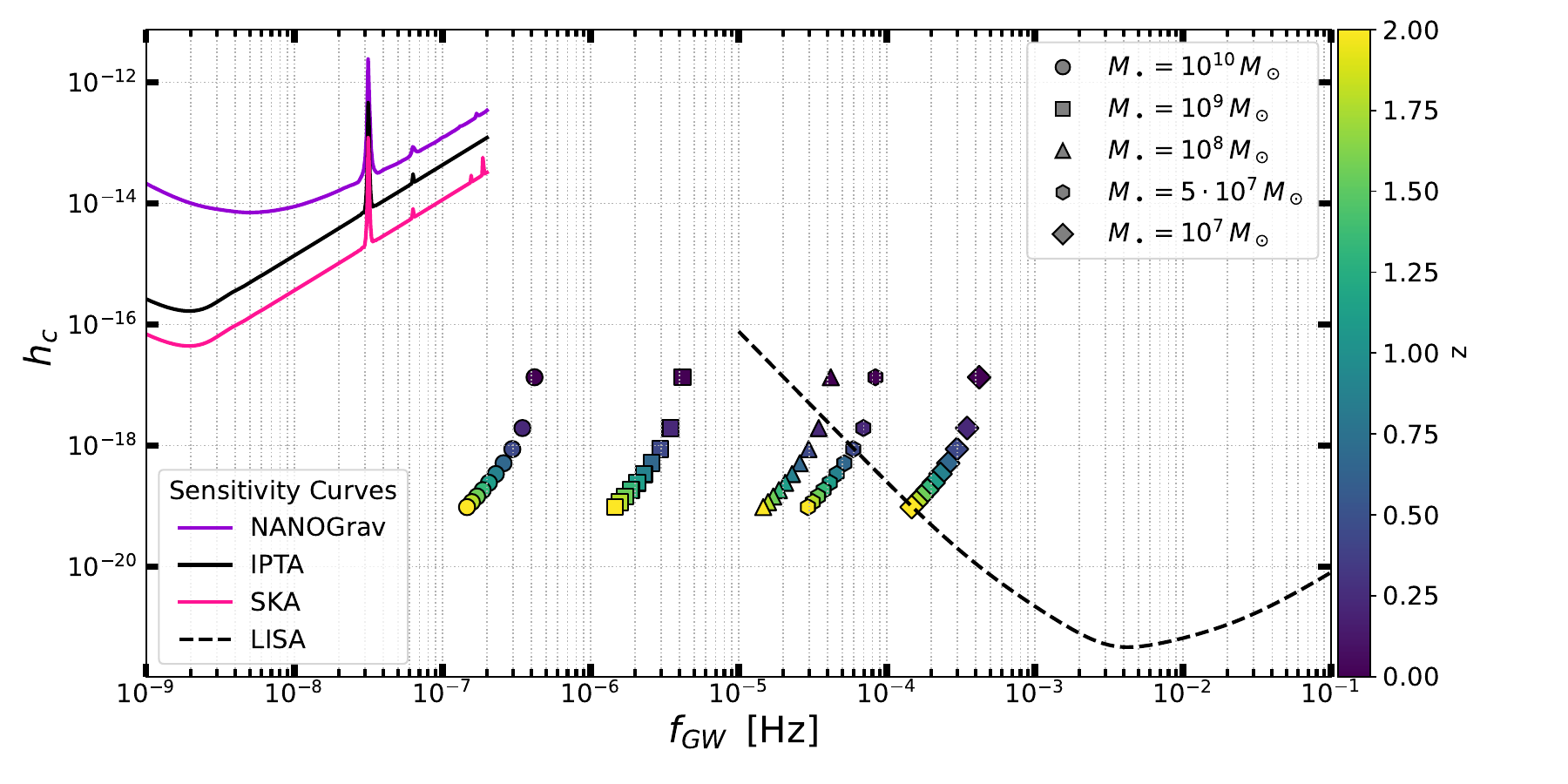}
    \caption{Characteristic strain for a range of heavy intermediate-mass-ratio inspiral (IMRI) sources. Different markers indicate varying primary SMBH masses and are color-coded by redshift in which we place the merging sources. Square markers represent our models, which lie just outside the LISA sensitivity band. Lower-mass primaries at modest redshifts can enter the LISA band, whereas higher-mass systems shift toward the PTA frequency range but remain below the detection threshold.}
    \label{fig:hc_fgw}
\end{figure*}

To compute the characteristic strain $h_{\mathrm{c,n}}$ of MBH mergers, we follow the formalism of \cite{Peters1963} and \cite{Peters1964}, which describe the gravitational-wave (GW) emission from eccentric binaries as a sum over harmonics $n$ of the orbital frequency $f_{\mathrm{orb}}$. The strain amplitude $h_{\mathrm{c,n}}$ for the $n-$th harmonic with frequency $f_{\mathrm{GW}}=nf_{\mathrm{orb}}$ $(n\geq1)$ is then given by \citep{Peters1963,Finn_2000,Barack_2004}

\begin{equation}
h_{c, n}^2=\frac{2}{3 \pi^{4 / 3}} \frac{G^{5 / 3}}{c^3} \frac{\mathcal{M}^{5 / 3}}{D_L^2} \frac{g(n, e)}{n^{2 / 3}} f_{\mathrm{orb}}^{-1 / 3},
\end{equation}

where $\mathcal{D}_{L}$ is the luminosity distance and $n=2$ circular orbits. Our goal is to assess whether these heavy IMRIs would be detectable by PTAs or LISA. To this end, we compute the characteristic strain $h_{\mathrm{c}}$ and GW frequency $f_{\mathrm{GW}}$ for circular binaries at $r=r_{\mathrm{isco}}$. In our models, the primary SMBH has a mass $M_\bullet=10^9 \: \mathrm{M_{\odot}}$ and the secondary $m=10^5 \: \mathrm{M_{\odot}}$, but we also explore how variations in the primary mass influence detectability. In addition, we examine the impact of different source redshifts on the prospects for detection.

Our IMRIs are indicated by the square markers in Fig.~\ref{fig:hc_fgw}. For these systems, the characteristic strain is too low for PTA detection, while the gravitational-wave frequency falls outside the LISA sensitivity band. We expect systems with more massive primaries to approach the PTA frequency range, whereas those with lower-mass primaries may enter the LISA band depending on their redshift. Notably, IMRIs around a primary of $M_\bullet=10^8 \: \mathrm{M_{\odot}}$ could become detectable if they are sufficiently nearby, suggesting that SMBH growth via such processes may be observed by future space-based interferometers such as LISA.

To obtain a more reliable assessment of the detectability of these sources with LISA, it is necessary to evaluate their signal-to-noise ratio (SNR), defined as \citep[e.g.,][]{Moore_2014,Robson_2019}. 

\begin{equation}
\label{eq:snr}
\mathrm{SNR}^2=\sum_n\left(\frac{h_{c, n}}{h_n\left(f_n\right)}\right)^2,
\end{equation}

where $h_n(f)=\sqrt{f S_n(f)}$ is the characteristic noise strain and $S_n(f)$ represents the one-sided power spectral density (PSD) of the detector's noise. For a typical LISA detection, we require $\mathrm{SNR}\geq8$ \citep[e.g.,][]{Amaro_Seoane_2017_LISA,Naoz_2023}. We note that there may exist a population of GW-driven mergers that is effectively unobservable in GWs with current or near-future detectors due to their low masses, large distances, or orbital parameters. For these systems, electromagnetic signatures may offer complementary probes of their existence \citep[e.g.,][]{D_Orazio_2023}.

The estimates for primaries with $M_\bullet = 10^7, 10^8$ and $10^{10}  \: \mathrm{M_{\odot}}$ shown in Fig.~\ref{fig:hc_fgw} are based on analytical estimates using Eqs.~\eqref{eq:f_peak}--\eqref{eq:snr}. These estimates assume circular binaries, a secondary mass of $m = 10^5 \mathrm{M_{\odot}}$, and, for illustrative purposes, are plotted at multiple redshifts to show how the same merger would appear when observed at different distances. By including these analytic points, we illustrate how the characteristic strain $h_{\mathrm{c}}$ and peak frequency scale with primary mass, highlighting the parameter space where future GW observatories may detect these heavy IMRIs. The sensitivity curve for LISA is taken from \cite{Robson_2019}, the NANOGrav curve from the public data release \citep{NANOGrav_2023}, and for SKA and IPTA we have used the Hasasia package \citep{Hazboun_2019}.

Binaries with primary masses of $M_\bullet = 10^7  \mathrm{M_{\odot}}$ can enter the LISA band out to redshifts as high as $z \sim 2$, although only systems with $z \lesssim 1.2$ reach a sufficiently large SNR for detection. For instance, a circular IMRI of this type would be observable at $z = 1.1$ with $\mathrm{SNR} = 8.62$, while the same system with a modest residual eccentricity of $e = 0.1$ could be detected out to $z = 1.35$ with $\mathrm{SNR} = 8.7$, underlining the critical role of residual eccentricity \citep{Fumagalli_2024, Garg_2024} not only as a tracer of the astrophysical environment but also as a factor enhancing source detectability. By contrast, circular IMRIs with primaries of $M_\bullet = 10^8 \mathrm{M_{\odot}}$ remain detectable only within $z \lesssim 0.04$ with $\mathrm{SNR} \geq 8$, increasing to $\mathrm{SNR} \sim 14$ for $e = 0.1$ at the same redshift. These examples highlight the strong potential of lower-mass primaries ($M_\bullet \sim 10^7$--$10^8 \mathrm{M_{\odot}}$) as promising hosts for heavy IMRIs detectable by LISA, motivating a systematic exploration of this regime in future work.

\section{Discussion and Conclusions}
\label{sec:Conclusions}

Massive early-type galaxies grow through a hierarchical assembly process, potentially leading to a population of low-mass MBHs as well as multiple high-mass SMBHs in their nuclei. This process takes place via major galaxy mergers ($M_1/M_2 \lesssim 4$), minor mergers ($4 \lesssim M_1/M_2 < 10$), very minor or mini mergers ($10 \lesssim M_1/M_2 < 100$), and also through accretion events ($M_1/M_2 > 100$), where the events with larger mass ratios are expected to be more common \citep[e.g.,][]{Nipoti_2025}. In this study we explore a scenario in which the central (primary) SMBH of the galaxy is accompanied by a population of lower-mass MBHs and study the dynamics of such systems with and without the presence of a companion SMBH. We find that the secondary SMBH can increase the total merger rate up to a factor of $\sim 5$, depending on the initial compactness of the lower-mass MBH system.

We set up initial conditions consisting of a small cluster of MBHs with mass $M_{\mathrm{MBH}}=10^5 \:  \mathrm{M_{\odot}}$ orbiting a central SMBH of mass $M_{\mathrm{SMBH}}=10^9 \: \mathrm{M_{\odot}}$. In the second set of simulations, we include an equal-mass companion SMBH on an outer binary orbit with semi-major axis $a_{b}=1 \: \mathrm{pc}$ and eccentricity $e_{\mathrm{b}}=0.5$, giving a pericentre distance of $r_{\mathrm{peri}}=0.5 \:\mathrm{pc}$. The MBHs are initialized within $r_{\mathrm{peri}}$, with semi-major axes drawn between $a_{\mathrm{min}}$ and $a_{\mathrm{max}}$, ensuring $a_{\mathrm{max}} \leq r_{\mathrm{peri}}$.  We performed simulations for three choices of the compactness of the semi-major axis distributions and for each case—both with and without the SMBH companion—we carry out 50 independent random realizations, giving a total of 300 runs.

For the simulations, we have used the direct-summation $N$-body integrator \mstar{} \citep{Rantala2020}, which enables the accurate modeling of non-softened dynamics via algorithmically regularised integration. The code incorporates relativistic effects with the inclusion of post-Newtonian corrections up to order 3.5PN and we have additionally implemented on-the-fly calculation of 3PN-accurate quasi-Keplerian elements following \cite{Memmesheimer_2004}. With the aim of classifying rapidly-evolving direct-plunge orbits on the primary SMBH versus the slowly-evolving GW-driven inspirals, we further implement an adaptive timestep method to properly capture (essential for a detailed output analysis) the latter type of merger orbits. 


To address the unphysical behavior of the 3PN orbital elements for binaries approaching merger \citep{Mannerkoski_2019}, we employ the geometric elements \citep{Csizmadia_2012}, which more reliably describe the late stages of slowly evolving orbits provided that the pericentre region is sufficiently well resolved. This requirement is met in our simulations through the use of the adaptive timestep scheme mentioned above. An alternative approach, which would avoid switching between different orbital element definitions and ensure full consistency with GW-driven evolution, would be to adopt a fully relativistic geodesic framework. In this case, the PN-corrected $N$-body integration would be terminated at a larger separation ($d \gg 10 R_{\mathrm{s}}$), and the final PN energy $E$ and angular momentum $L$ extracted and mapped onto the semi-latus rectum $p$ and eccentricity $e$ \citep[e.g.,][]{Hughes_2024}. The resulting $(p,e)$ pair could then be used to compute and evolve the corresponding geodesic \citep[e.g.,][]{Gair_2006}, enabling a physically motivated discrimination between GW-driven inspirals and true dynamical plunges through the evolution of the constants of motion. This methodology has recently been applied by \citet{Mancieri_2025a} to infer EMRI eccentricity distributions and obtain robust event rates relevant for LISA, and we are currently pursuing a similar approach as part of a follow-up work.

The orbital and dynamical evolution of the lower-mass MBH system is dictated by the competition between various dynamical processes, like resonant and non-resonant relaxation (RR and NR), relativistic precession, von-Zeipel-Kozai-Lidov oscillations and the energy and angular momentum dissipation due to the emission of gravitational waves. In all systems considered, the characteristic timescale of GR precession is many orders of magnitude shorter than that of RR, implying that RR is effectively quenched by relativistic effects \citep{Merritt_2011,Brem_2014,Hamers_2014}. The presence of a companion SMBH can induce ZKL oscillations, though these are also largely suppressed by GR \citep{Holman_1997,Bonetti_2016,Lim_2020}, except in the case of the widest MBH configuration, where the secondary SMBH disrupts the system and produces an enhanced rate of DPs or a burst of ejections within the first $t\lesssim0.6 \: \mathrm{Myr}$. In contrast, the uncorrelated nature of gravitational encounters from NR are present, competing with GR and shaping the long-term evolution of the system. Diffusion of angular momentum through relaxation processes can gradually push MBHs toward higher eccentricities, while strong chaotic encounters provide a more violent pathway, rapidly scattering MBHs onto nearly radial orbits. Together, these mechanisms drive a substantial fraction of MBHs into highly eccentric trajectories, leading to a large number of DP merger orbits. We note, however, that phenomena of this kind could potentially be explained within a fractional-dynamics framework, in which angular-momentum transport is intrinsically non-local, allowing for sporadic \emph{L\'evy} jumps that can effectively bypass the Schwarzschild barrier \citep{Amaro_Seoane_2025_Fractional}. Exploring whether the \emph{elusive} plunges observed here are manifestations of such non-local diffusion lies beyond the scope of the present work and is deferred to future studies.

In our analysis we focus on the SMBH-MBH mergers involving the primary SMBH. Only four mergers with the secondary SMBH occurred in the entire sample of $300$ simulations. No MBH-MBH mergers occurred in our simulation set. Our simulations demonstrate that the presence of a secondary SMBH substantially enhances the overall merger rate of MBHs, particularly for wide initial semi-major axis distributions. In these cases, the number of mergers increases by a factor of $\sim 5$ compared to the single-SMBH runs. Even for more compact initial configurations, the enhancement remains significant, with merger rates elevated by a factor of $\sim 2$. This strong boost in the SMBH-MBH merger efficiency highlights the importance of binary-SMBH environments for the evolution of central MBH subsystems.

The character of these mergers differs markedly between single- and binary-SMBH systems as presented in Figure~\ref{fig:merger_pie}. For wide configurations, the inspiral fraction is significantly higher in the single-SMBH case, reaching $50\%$ for $a_{\mathrm{min}} = 0.05 \: \mathrm{pc}$ and $29.6\%$ for $a_{\mathrm{min}} = 0.01 \: \mathrm{pc}$. In contrast, binary-SMBH runs produce only $2.3\%$ and $6.8\%$ inspirals for the same configurations. The presence of the secondary SMBH efficiently drives MBHs onto nearly radial trajectories, suppressing gradual inspirals and instead producing rapid plunges. In the most compact configuration ($a_{\mathrm{min}} = 0.005\: \mathrm{pc}$), the inspiral fractions become more balanced, with $20.7\%$ in the binary-SMBH case compared to $33.3\%$ for the single-SMBH case. This indicates that relaxation processes dominate inspiral formation in isolated systems, while binary environments preferentially produce plunges. These findings contrast with studies of lower-mass systems (primary $M_{\bullet} \sim 10^6  M_\odot$), where secondary SMBHs enhance EMRI production \citep{Naoz_2022,Mazzolari_2022}.

An additional outcome concerns the production of hyperbolic plunges and high-velocity ejections. For the two widest configurations, plunge fractions in the binary-SMBH case reach $93.2\%$ and $97.7\%$ of the total mergers, with nearly $\sim 30\%$ of plunges in the widest setup (Fig.~\ref{fig:merger_pie}) occurring with $e \geq 1$, corresponding to hyperbolic encounters where MBHs are dynamically scattered and directly driven onto the central SMBH. The secondary SMBH also acts as an efficient channel for ejecting MBHs via the slingshot mechanism \citep{Hills_1988} from the system: in the wide configuration, 283 out of 419 MBHs are ejected from the host galaxy within the first $\sim 0.5 \: \mathrm{Myr}$, coinciding with the early pericentre passages of the secondary. Intermediate and compact systems similarly show enhanced ejection numbers compared to single-SMBH runs, from 12 (out of 19) to 44 (out of 64) for the intermediate configuration, and from 15 (out of 24) to 27 (out of 36) for the compact one.

On average out of $10$ MBHs, the single-SMBH systems yield fewer than one escaper per realization across all initial configurations, whereas the binary-SMBH cases produce markedly higher ejection rates: $\sim 8$ escapers per run in the wide setup, $\sim 1.3$ in the intermediate one, and $\sim 0.7$ in the compact case. The reduced ejection numbers in more compact systems indicate that the efficiency of the slingshot mechanism is suppressed when the MBHs are more tightly bound, further highlighting the dominant role of the secondary SMBH in driving high-velocity ejections in wider configurations. Most ejections occur with velocities of a few $10^3 \: \mathrm{km \: s^{-1}}$, though in rare compact cases velocities exceed $10^4 \: \mathrm{km \: s^{-1}}$. These results demonstrate the efficiency of hard MBH binaries in producing high-velocity ejections, in agreement with previous studies of both SMBH–IMBH and MBH–MBH binaries in galactic nuclei and star clusters \citep{Quinlan_1996,Baumgardt_2006,Sesana_2006,Sesana_2007,Rantala2018,Rasskazov_2019,Evans_2023,Souvaitzis_2025}.

The eccentricity distribution of merger events encodes key information about their dynamical environment. High residual eccentricities ($e \gtrsim 0.9$) can signal triple-MBH interactions or chaotic multi-body encounters, accelerating coalescence and requiring accurate relativistic treatment \citep[e.g.,][]{Bonetti_2016}. In our simulations, most direct plunges occur at $e \geq 0.9$, while inspirals generally have $e < 0.5$, with a subset of IMRIs retaining $e \gtrsim 0.6$ due to chaotic encounters in compact configurations. Such high eccentricities leave distinct signatures on GW waveforms, which are particularly relevant for LISA \citep{Bonetti_2019}. Moreover, the residual eccentricity encodes information about the surrounding stellar \citep{Matsubayashi_2007,Lockmann_2008,Preto_2009,Sesana_2010,Gualandris_2022} or gaseous \citep{Sesana_2005,MacFadyen_2008,Cuadra_2009,Roedig_2012,Siwek_2023,Tiede_2024} environment, producing measurable modifications in the GW signal \citep{Garg_2024} that provide a unique probe of MBH assembly channels.

The characteristic chirp mass of our simulated binaries, with a primary SMBH of $M_\bullet = 10^9 \: \mathrm{M_\odot}$ and a secondary MBH of $m = 10^5 \: \mathrm{M_\odot}$, is $\mathcal{M} \simeq 4 \times 10^6 \: \mathrm{M_\odot}$, placing them within the nominal LISA sensitivity range $\mathcal{M} \sim 10^4$–$10^7 \: \mathrm{M_\odot}$ \citep{Klein_2016,Amaro_Seoane_2017_LISA}, but below the typical PTA sensitivity threshold $\mathcal{M} \sim 10^8$–$10^{10} \: \mathrm{M_\odot}$ \citep{Hobbs_2010,Sesana_2013}. Despite falling within LISA’s chirp-mass range, the observed GW frequencies of our systems are generally shifted outside LISA’s most sensitive band, mainly due to the large primary SMBH masses, while the characteristic strain $h_{\mathrm{c}}$ of our IMRIs is too low for PTA detection. A more refined assessment of detectability would require estimating the number of GW cycles accumulated in the LISA band by evolving the system adiabatically as a sequence of geodesic orbits, with the constants of motion updated using orbit-averaged fluxes. Such a treatment, which goes beyond the plunge–inspiral classification used here, is left to future work. For example, IMRIs around a primary of $M_\bullet = 10^9 \: \mathrm{M_\odot}$ largely remain undetectable, while lower-mass primaries ($M_\bullet \sim 10^7$–$10^8 \: \mathrm{M_\odot}$) can reach the LISA band at sufficiently low redshifts, illustrating the importance of both mass and distance in setting their GW detectability. These considerations also suggest that a population of mergers may remain essentially “invisible” to current GW observatories, motivating complementary electromagnetic searches \citep[e.g.,][]{D_Orazio_2023}.

To provide a broader perspective on detectability, we complement our simulations with analytical estimates of binaries with primary masses $M_\bullet = 10^7, 10^8,$ and $10^{10} \: \mathrm{M_\odot}$, assuming circular binaries, $m = 10^5 \: \mathrm{M_\odot}$, and plotting them at multiple redshifts to illustrate how the same merger would appear at different distances (Fig.~\ref{fig:hc_fgw}). For instance, a circular IMRI with $M_\bullet = 10^7 \: \mathrm{M_\odot}$ would be detectable by LISA at $z = 1.1$ with $\mathrm{SNR} = 8.62$, while a modest eccentricity of $e = 0.1$ increases the detectable horizon to $z = 1.35$ with $\mathrm{SNR} = 8.7$ \citep{Fumagalli_2024, Garg_2024}. By contrast, IMRIs with $M_\bullet = 10^8 \: \mathrm{M_\odot}$ are detectable only within $z \lesssim 0.04$, with $\mathrm{SNR} \sim 14$ if $e = 0.1$. These results underscore the strong potential of lower-mass primaries as hosts for LISA-detectable IMRIs and motivate a systematic follow-up study to explore their occurrence rates, residual eccentricities, and observational signatures across a broader black hole mass range.

\section*{Acknowledgements}

L.S. would like to thank Davide Mancieri, Matteo Bonetti and Jacob Stegmann for insightful discussions and for their valuable comments. We thank the anonymous referee for constructive feedback that helped improve the clarity and quality of this manuscript.
TN acknowledges the support of the Deutsche Forschungsgemeinschaft (DFG, German Research Foundation) under Germany’s Excellence Strategy - EXC-2094 - 390783311 of the DFG Cluster of Excellence ''ORIGINS''.

\section*{SOFTWARE}

\mstar{} \citep{Rantala2020}, NumPy \citep{Harris_2020}, SciPy \citep{Virtanen_2020}, Matplotlib \citep{Hunter_2007}, Hasasia \citep{Hazboun_2019}.

\section*{Data Availability}

The data underlying this article will be shared on reasonable request
to the corresponding author.


\bibliographystyle{mnras}
\bibliography{main} 

\appendix
\section{Poisson Confidence Intervals for Merger Fractions}
\label{apndx:A}

To estimate uncertainties on the merger fractions reported in this work, we adopt the analytical approximation introduced by \cite{Gehrels_1986}, which is specifically designed for Poisson-distributed event counts in the low-number regime. In our simulations, each configuration consists of a finite number of realizations, and several merger channels—most notably hyperbolic plunges—are rare, with observed counts as low as $N = 0$ or $N = 1$. In such cases, standard Gaussian approximations (e.g. $\sigma = \sqrt{N}$) are known to fail, as they do not capture the intrinsic asymmetry of the Poisson distribution and can substantially underestimate the true confidence bounds.

Gehrels derived closed-form expressions that approximate the exact Poisson confidence intervals to within $1$–$2\%$ accuracy over the full range of event counts, including the extreme low-count limit. For a one-sigma ($84\%$ confidence level) interval, the upper and lower limits on the true mean event rate $\lambda$ for an observed count $N$ are given by

\begin{equation}
\lambda_{\mathrm{up}} \simeq (N+1)\left(1 - \frac{1}{9(N+1)} + \frac{1}{3\sqrt{N+1}}\right)^3 ,
\end{equation}
\begin{equation}
\lambda_{\mathrm{low}} \simeq N\left(1 - \frac{1}{9N} - \frac{1}{3\sqrt{N}}\right)^3 .
\end{equation}

\begin{figure*}
    \includegraphics[width=1\textwidth, keepaspectratio]{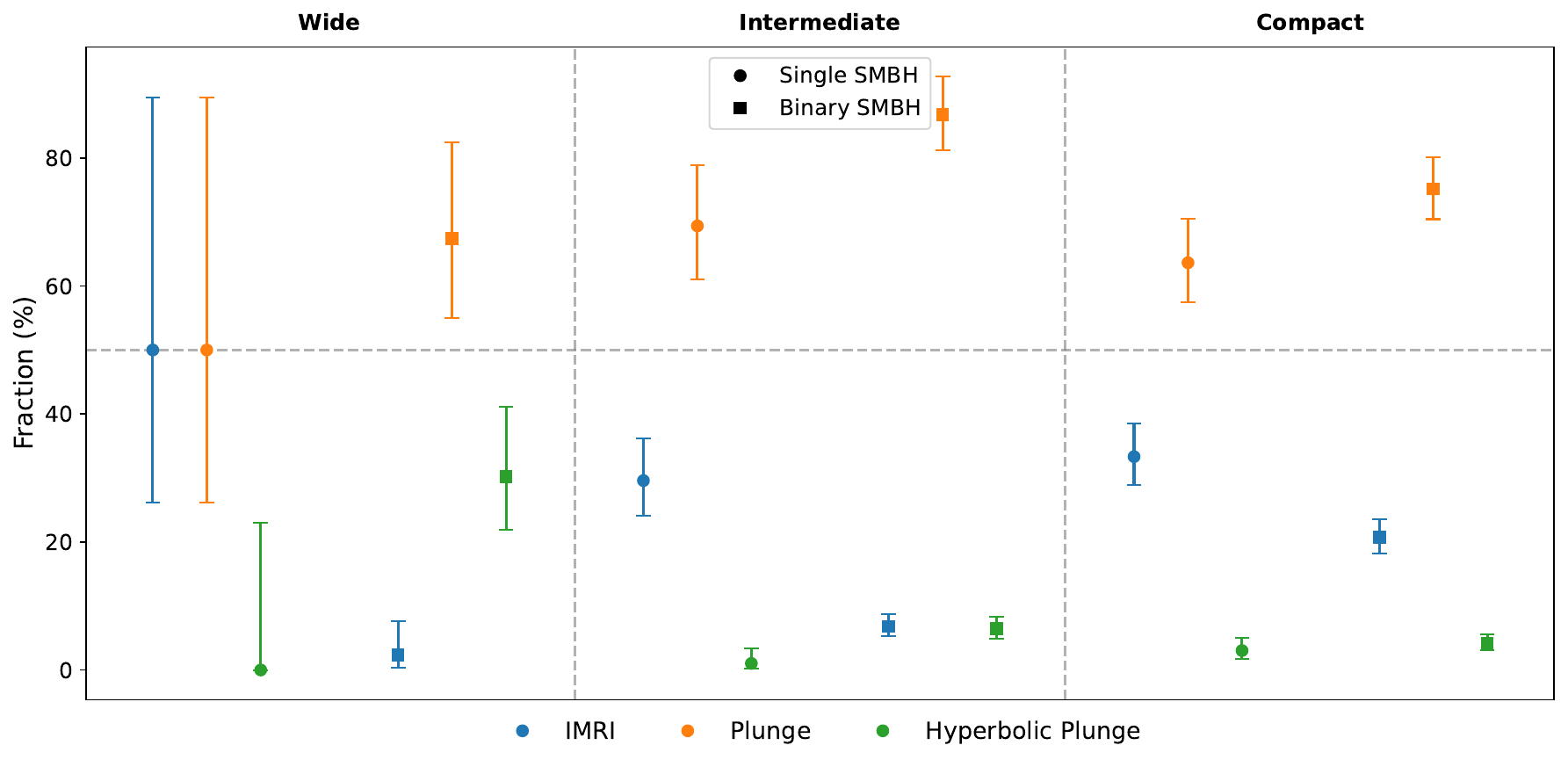}
    \caption{Merger fractions for different initial semi-major axis distributions, shown separately for single-SMBH (circles) and binary-SMBH (squares) systems. Colors indicate the merger channel: IMRIs (blue), direct plunges (orange), and hyperbolic plunges (green). The uncertainty estimates (error bars here) represent asymmetric 1$\sigma$ Gehrels confidence intervals appropriate for Poisson-distributed event counts. Vertical dashed lines separate the wide, intermediate, and compact initial configurations. The horizontal dashed line marks a $50\%$ fraction for reference. Large uncertainties for rare channels, particularly merger events in the wide configuration with single-SMBH runs, reflect the small-number statistics inherent to these events.}
    \label{fig:A1}
\end{figure*}

For $N=0$, the lower bound is set to $\lambda_{\mathrm{low}} = 0$, while the upper bound remains finite. These limits are computed for each merger channel and converted into asymmetric confidence intervals on the corresponding merger fractions by normalizing to the total number of realizations in each configuration. As expected, the resulting uncertainties become large when the number of detected events is small, reflecting the intrinsic statistical limitations of rare-event measurements.

The Gehrels confidence intervals described above are applied to all merger fractions shown in Fig.~\ref{fig:A1}. For each initial configuration and system type (single or binary SMBH), the number of events in each merger channel (IMRIs, direct plunges, and hyperbolic plunges) is treated as a Poisson count, and the corresponding asymmetric 1$\sigma$ confidence limits are computed using Eqs.~(A1)–(A2). These limits are then normalized by the total number of realizations in each bin to obtain the uncertainty estimates on the reported fractions. The resulting uncertainties are particularly large for rare outcomes—most notably for merger events in the wide configuration with single-SMBH runs—reflecting the low number of detected events in these channels. Conversely, merger channels with large event counts exhibit comparatively smaller relative uncertainties, demonstrating that the observed trends across different initial separations are statistically robust despite the low-$N$ regime. In particular, the systematic dominance of direct plunges in binary-SMBH configurations persists well outside the estimated confidence intervals, indicating that this behavior is not driven by statistical fluctuations.


\bsp	
\label{lastpage}
\end{document}